\documentclass[aps,prd,onecolumn,nofootinbib,notitlepage,showkeys,superscriptaddress,preprintnumbers]{revtex4-2}

\usepackage{graphicx}
\usepackage{dcolumn}
\usepackage{bm}

\usepackage{amsmath,amssymb}
\usepackage{float}
\usepackage{multirow}
\usepackage{slashed}
\usepackage{xcolor}
\usepackage{physics}
\usepackage{multirow}
\usepackage[colorlinks=true, pdfstartview=FitV, bookmarks=true, bookmarksnumbered=true, breaklinks]{hyperref}

\usepackage{mathtools,braket}
\usepackage{soul}
\usepackage{lipsum}  
\usepackage{color}
\definecolor{blue}{rgb}{0.0, 0.0, 1.0}
\definecolor{red}{rgb}{1.0, 0.0, 0.0}
\definecolor{royalblue}{rgb}{0.0, 0.14, 0.4}
\hypersetup{linkcolor=royalblue, citecolor=blue, urlcolor=royalblue}

\usepackage{hyperref}
\hypersetup{colorlinks=true,citecolor=blue,linkcolor=blue,urlcolor=blue}

\usepackage[mathlines]{lineno}

\usepackage{newtxmath}
\usepackage{mathrsfs}

\usepackage[normalem]{ulem}
\usepackage{color}

\usepackage{tikz,xcolor,hyperref}
\definecolor{lime}{HTML}{A6CE39}
\DeclareRobustCommand{\orcidicon}{%
	\begin{tikzpicture}
	\draw[lime, fill=lime] (0,0) 
	circle [radius=0.16] 
	node[white] {{\fontfamily{qag}\selectfont \tiny ID}};
	\draw[white, fill=white] (-0.0625,0.095) 
	circle [radius=0.007];
	\end{tikzpicture}
	\hspace{-2mm}
}

\foreach \x in {A, ..., Z}{%
	\expandafter\xdef\csname orcid\x\endcsname{\noexpand\href{https://orcid.org/\csname orcidauthor\x\endcsname}{\noexpand\orcidicon}}
}

\begin{document}
\preprint{LFTC-25-08/102}
\title{Medium effects on the electromagnetic form factors of the $\rho$ meson}

\author{Parada T.~P.~Hutauruk\orcidA{}} 
\email[E-mail: ]{phutauruk@gmail.com}
\affiliation{Department of Physics, Pukyong National University (PKNU), Busan 48513, Korea}
\affiliation{Departemen Fisika, FMIPA, Universitas Indonesia, Depok 16424, Indonesia}
\author{Terry Mart\orcidC{}}
\email[E-mail: ]{terry.mart@sci.ui.ac.id}
\affiliation{Departemen Fisika, FMIPA, Universitas Indonesia, Depok 16424, Indonesia}
\author{Kazuo Tsushima\orcidB{}}
\email[E-mail: ]{kazuo.tsushima@gmail.com, kazuo.tsushima@cruzeirodosul.edu.br} %
\affiliation{Laboratório de Física Teórica e Computacional-LFTC, Programa de P\'{o}sgradua\c{c}\~{a}o em Astrof\'{i}sica e F\'{i}sica Computacional,
Universidade Cidade de S\~{a}o Paulo, 01506-000 S\~{a}o Paulo, SP, Brazil}\textbf{}
\date{\today}

\begin{abstract}
The dynamics of partons inside the light $\rho$ meson is found to be essential for its
properties and internal structure, both in free space and in the nuclear medium.
In this paper, we systematically investigate the in-medium structure changes of $\rho^+$ mesons within
the covariant Nambu-Jona-Lasinio (NJL) model, utilizing the Schwinger proper-time
regularization scheme. We solve the Bethe-Salpeter equations to guarantee the bound meson-state
condition. At the quark level, the nuclear medium effects are also derived within the same NJL model to maintain a consistent approach with the in-medium
$\rho^+$ meson electromagnetic form factors. To this end, we analyze the spacelike
electromagnetic form factors of the $\rho^+$ meson in free space and in a nuclear
medium. We find that the charge radius and quadrupole moment of the $\rho^+$ meson increase
with increasing nuclear matter density, while the magnetic moment decreases, in agreement
with the existing previous theoretical predictions.
The enhancement of the $\rho^+$ meson charge radius at
normal density relative to that in free space is about 11\% (0.08 fm), while the reduction of
$\rho^+$ meson magnetic moment is about 8\% (0.20 $\mu_N$). Our predictions for the charge
radius, magnetic moment, and quadrupole moment of the $\rho^+$ meson in both free space and
nuclear medium, remain challenging to be verified experimentally.
\end{abstract}
\maketitle

\section{Introduction} \label{sec:intro}
Quantum chromodynamics (QCD) is widely believed to be an underlying theory of strong
interaction of the Standard Model (SM)~\cite{Ioffe:2005ym}. The nonperturbative
QCD at the low-energy scale depends sensitively on the strong running coupling
constant $\alpha_s(Q^2)$. Recent interest in the running coupling has increased in focusing on its
behavior in relation to gluon saturation phenomena at small quark longitudinal momentum
$x$ (Bjorken-$x$) and the low negative values of the
squared four-momentum transfer ($q^2 \equiv - Q^2 < 0$),
where nonlinear QCD effects become significant~\cite{Deur:2016tte,Kovchegov:2023bvy}.
Rather different from the perturbative QCD aspects, which are relatively well understood and more
straightforward to apply, the nonperturbative QCD, which has the properties of confinement and spontaneous or dynamical chiral symmetry breaking (SCSB or DCSB), is still not fully understood until now.
These features of nonperturbative QCD are expected to be better understood by studying the
internal structure of hadrons, such as the parton distribution functions (PDFs),
electromagnetic form factors (EMFFs), parton distribution amplitudes (PDAs),
transverse momentum dependent parton distribution functions (TMD-PDFs), gravitational form
factors (GFFs), and generalized parton distribution functions (GPDs), in free space as well as
in a nuclear medium. As is well known, hadrons can
be described effectively as either three
valence dressed quarks for baryons or a dressed valence quark-antiquark pair for mesons,
with their internal quark dynamics governed by the strong interaction of QCD.

In the 50 years of the existence of QCD~\cite{Gross:2022hyw}, much impressive progress has 
been made in studying the free space properties of the hadrons to better understand
QCD~\cite{Braguta:2004kx,Cardarelli:1994yq,deMelo:2012hj,Jaus:2002sv,Aliev:2009gj,
Bhagwat:2006pu,Xu:2024fun,Shi:2023oll,Carrillo-Serrano:2015uca,Qian:2020utg,Haberzettl:2019qpa,
Krutov:2018mbu}. The information on the internal structure of hadrons has been obtained through the
PDFs~\cite{Liu:2023fpj,Kumano:2021xau}, EMFFs~\cite{Carrillo-Serrano:2015uca},
PDAs~\cite{Gao:2014bca,Ball:2007zt,Arifi:2023jfe}, TMDs~\cite{Kumano:2020ijt,Ninomiya:2017ggn},
GFFs~\cite{Sun:2020wfo,Kim:2022wkc}, and
GPDs~\cite{Sun:2020jng}. These physical properties have been extensively studied, mostly in
free space. In addition to the studies in free space, the features of QCD can also be explored by
investigating hadrons in a medium, e.g., through their structure functions in nuclei or nuclear
matter~\cite{Geesaman:1995yd}.
The European Muon Collaboration (EMC) experiment provided clear evidence that the hadron unpolarized structure function is modified in a nuclear medium.
This modification, known as the \textit{EMC effect}, is
reflected in the ratio of the unpolarized structure function of a bound hadron to that in
free space showing less than unity~\cite{EuropeanMuon:1983wih}, in particular at the
large Bjorken-$x$ region.
Since then, studies of hadron structure and properties in the nuclear medium have expanded
significantly (see Refs.~\cite{Geesaman:1995yd,Hayano:2008vn,Hen:2016kwk} and references therein).
This has primarily motivated us to study the $\rho^+$ meson EMFFs in the nuclear medium, which
constitutes the central focus of the present work.

The $\rho$ mesons have interesting internal quark dynamic features compared to those of the
pions. For instance, even though both $\pi^+$ and $\rho^+$ mesons consist of
the $u$ and $\bar{d}$
light dressed quarks and antiquarks, their physical masses, and internal structures are much different because of the spin interactions of the dressed quark and antiquark contents inside the mesons.
This difference exhibits that the $\rho$ meson with mass
$m_\rho = 0.770$ GeV has spin$-1$ and odd parity $J^P = 1^-$, and the pion with mass
$m_\pi = 0.140$ GeV has spin$-0$ and odd parity $J^P =0^-$.
Delving deeper into the $\rho$ meson internal structure will lead us to a better understanding of the contribution of quarks or quark dynamics to
its internal structure, and one can gain new
insight into physics features based on QCD. So far, to gain a better understanding of the role of the nuclear medium on hadron structure and QCD properties in the medium, several studies have been performed
using different theoretical models and approaches in
Refs.~\cite{Arifi:2023jfe,Gifari:2024ssz,Hutauruk:2018qku,Hutauruk:2016sug,
Hutauruk:2021kej,Farrar:1979aw,Bernard:1988bx,Hutauruk:2019ipp},
in particular for the pion EMFFs. This is supported
by the availability of experimental data for the free space pion EMFFs,
even for only the intermediate $Q^2$ range, where $(Q^2 = -q^2 > 0)$ is the negative of
squared four-momentum transfer. However, experimental (empirical) data for the $\rho$ meson
EMFFs in free space or in the nuclear medium are still scarce compared to those for the pion.
In the past, several experiments have been performed at the CERN SPS through the
CERES/NA45 collaboration~\cite{Cassing:1995zv}, CLAS g7 at JLAB~\cite{CLAS:2008jqp,CLAS:2007dll},
and KEK Proton-Synchrotron~\cite{Naruki:2005kd,E325:2000smw} to measure or extract the $\rho$
meson properties in the nuclear medium, such as the mass, the spectral density function, and the
decay width.
Besides the experimental efforts, extensive theoretical studies have been carried out to investigate
the properties of $\rho$ mesons using various models and approaches, such as QCD sum rule
(QCDSR)~\cite{Hatsuda:1991ez,Asakawa:1993pq}, inverse QCD sum rule (IQCDSR)~\cite{Mutuk:2025lak},
chiral perturbation theory (ChPT)~\cite{Cabrera:2000dx}, generalized Nambu--Jona-Lasinio (GNJL)
model~\cite{Bernard:1988db}, vector meson dominance (VMD)
model~\cite{Ko:1994hc,Chanfray:1992kp,Rapp:1997fs}, and dispersion
relation~\cite{Kondratyuk:1998ec}. These studies have been performed mostly on the
in-medium $\rho$ meson properties focusing on the chiral symmetry
restoration, while a study of the $\rho$ meson EMFFs in the nuclear medium is still scarce in the literature.

Recently, investigations of the free-space $\rho$-meson properties, such as its electromagnetic form factors (EMFFs), including the charge radius, magnetic moment, and quadrupole moment, have been extended in order to study their behavior in the nuclear medium~\cite{deMelo:2018hfw}.
The authors of Ref.~\cite{deMelo:2018hfw} reported
interesting features of the $\rho$ meson EMFFs in the nuclear medium, which were calculated in a
hybrid model; combining the light front constituent quark model (LFCQM) and quark-meson
coupling (QMC) model.
They found that the quadrupole form factor, $G_Q(Q^2)$, exhibits intriguing behavior in the
nuclear medium. In particular, at low momentum transfer ($Q^2 \simeq 3$ GeV$^2$), the values of
$G_Q(Q^2)$ at $\rho_B/\rho_0 = 0.25$ and $0.50$ intersect with the free-space result
at $\rho_B/\rho_0 = 0.0$. This notable feature has motivated us to further investigate the $\rho^+$
meson EMFFs in the nuclear medium to better understand the underlying origin and implications of the behavior in
the quadrupole form factor.

In the present study, using a consistent treatment approach, we systematically investigate
the $\rho^+$ meson EMFFs incorporating the charge/electric
$G_C^* (Q^2)$, the magnetic moment $G_M^*(Q^2)$, and the quadrupole moment $G_Q^* (Q^2)$
in symmetric nuclear matter within the framework of the covariant NJL model
using the Schwinger proper time regularization scheme-simulating the quark confinement, where the
medium effect is also calculated within the NJL model at the quark level~\cite{Bentz:2001vc}.
The NJL model has been successfully employed in describing the hadron internal structures in free
space and nuclear medium, e.g., the $\rho$ meson
EMFFs~\cite{Carrillo-Serrano:2015uca,Cloet:2014rja}, PDF~\cite{Hutauruk:2016sug}, and $\rho$ meson
TMD-PDFs~\cite{Ninomiya:2017ggn}. In this work,
we compute the $\rho^+$ meson effective mass, in-medium $\rho$ meson-quark coupling constant, and in-medium modifications of the $\rho^+$ meson EMFFs: charge radius, magnetic moment, and
quadrupole moment in the NJL model. As a result, we find that the $\rho$ meson effective mass
decreases as the nuclear matter density increases. The reduction in the $\rho$ meson mass at
$\rho_B/\rho_0 =1.0$  with $\rho_0 = 0.16$ fm$^{-3}$ relative to that in free space
is about 10\%, which is consistent with that obtained by the QCDSR in Ref.~\cite{Hatsuda:1991ez},
and very recent analysis using the IQCDSR in Ref.~\cite{Mutuk:2025lak}, which are approximated by
(10-20)\%. However, our result in the $\rho$ mass reduction is rather different from that obtained
in the hybrid model of Ref.~\cite{deMelo:2018hfw}, where they reported the reduction of about 33\%. For the $\rho$ meson-quark coupling constant, we find that it decreases as the nuclear
matter density increases.
In this study, we also find that the in-medium charge radius and quadrupole moment form factors of the $\rho^+$ meson increase as the nuclear matter density increases,
while the in-medium magnetic moment form factor decreases.

This paper is organized as follows.
In Sec.~\ref{sec:vacuumNJL}, we briefly describe the free space $\rho$ meson properties and
structure in the covariant NJL model with the help of the Schwinger proper-time regularization
scheme to simulate the quark confinement, mimicking QCD properties. Section~\ref{sec:NMNJL}
describes the symmetric nuclear matter in the NJL model and the calculations of the light
quark effective masses. We then present the mathematical expression of the $\rho$ meson properties
in the nuclear medium in Sec.~\ref{sec:vectormediumnNJL}, while a formulation of the $\rho$ meson
EMFFs in the nuclear medium is presented in Sec.~\ref{sec:ffv}. Section~\ref{sec:MR} presents the
numerical results for the $\rho$ meson properties in the nuclear medium and free space. A summary
and conclusion are devoted to Sec .~\ref{sec:summary}.

\section{Free space $\rho$ meson properties} 
\label{sec:vacuumNJL}
In this section, we present the description of the SU(2) flavor NJL model effective Lagrangian,
in addition to the free space quark and $\rho$ meson properties in the NJL model.

\subsection{SU(2) NJL Lagrangian}
The NJL model is a Poincar\'{e}-covariant
quantum field theory-based model that shares QCD's main
properties at low energy scales. For instance, the NJL model encapsulates the emergent
features of QCD, such as the dynamical chiral symmetry breaking and confinement in some versions, like the presently used one. The dynamical
chiral symmetry is clearly shown in the NJL model through the chiral condensate in the NJL gap
equation, as the order parameter of the chiral symmetry breaking, which dynamically generates
dressed quark mass from the bare (current) quark mass. As the original NJL model is not a
confining model, the confinement in the present study is described through the Schwinger
proper-time regularization scheme,
which completely removes the unphysical threshold that allows the hadron decay into quarks.
An SU(2) flavor symmetric NJL model effective Lagrangian can be written in terms of the four-fermion
contact interactions,
where the gluons are integrated out and absorbed into the coupling constants with dimension,
given by~\cite{Klevansky:1992qe}
\begin{eqnarray}
    \label{eq:vacNJL1}
    \mathscr{L}_{\mathrm{NJL}} &=& \bar{\psi}_q \big( i \partial \!\!\!/ - \hat{m}_q \big) \psi_q +
\frac{1}{2}
    G_\pi \Big[ \big( \bar{\psi}_q \psi_q \big)^2 - \big( \bar{\psi}_q \vec{\tau} \gamma_5 \psi_q
\big)^2 \Big] - \frac{1}{2} G_\omega \big( \bar{\psi}_q \gamma^\mu \psi_q \big)^2 - \frac{1}{2}
G_\rho \Big[ \big( \bar{\psi}_q \gamma^\mu \vec{\tau} \psi_q \big)^2 + \big( \hat{\psi}_q
 \gamma^\mu \gamma_5 \vec{\tau}\psi_q \big)^2 \Big],
\end{eqnarray}
where the $\psi_q^{T} = \big( \psi_u, \psi_d\big)$ represents the quark field with flavor
$q = (u,d)$, $
\vec{\tau}$ is the Pauli matrices, and $\hat{m}_q =\mathrm{diag} (m_u, m_d)$ is the current quark
mass matrix. It is worth noting that one group of the four-fermion interaction terms is
proportional to the coupling constant $G_\pi$, which governs the strength of the direct terms of the antiquark-quark
interaction in the scalar and pseudoscalar channels, and is responsible for the dynamical chiral
symmetry breaking (DCSB). The constants $G_\omega$ and $G_\rho$ turn out to represent
the $\omega$ and $\rho$ meson coupling constants, respectively.

Using the mean-field approximation, we evaluate the dressed quark masses through the quark self-energy
interaction. The gap equation in the Schwinger proper-time regularization scheme can be written as
\begin{eqnarray}
    \label{eq:vacuumNJL2}
    M_q &=& m_q + 12i G_\pi \int \frac{d^4k}{(2\pi)^4} \mathrm{Tr} \big[ S_q (k) \big], \nonumber \\
     &=& m_q - 4 G_\pi \big< \bar{\psi}_q \psi_q \big> = m_q + \frac{3G_\pi M_q}{\pi^2}
     \int_{\tau_{\mathrm{UV}}}^{\tau_{\mathrm{IR}}} \frac{d\tau}{\tau^2} e^{-\tau M_q^2},
\end{eqnarray}
where $\tau_{\mathrm{UV}} = \Lambda_{\mathrm{UV}}^{-2}$ and $\tau_{\mathrm{IR}} =
\Lambda_{\mathrm{IR}}^{-2}$
denote the ultraviolet (UV) and infrared (IR) integration limits, respectively. The IR value
is fixed at $\Lambda_{\mathrm{IR}} = 0.24~\text{GeV}$, corresponding to $\Lambda_{\mathrm{QCD}}$,
while the UV value $\Lambda_{\mathrm{UV}}$ is determined by fitting the pion mass ($m_\pi =
0.14~\text{GeV}$) and decay constant ($f_\pi = 0.093~\text{GeV}$). In other words,
$\Lambda_{\mathrm{UV}}$ and $\Lambda_{\mathrm{IR}}$ are known as the regularization scales (regulators) of
the NJL model. $M_q$ is the (dynamical) dressed quark mass for quark flavor $q$,
and the trace, "$\mathrm{Tr}$", appearing in the above, operates only over the Dirac
indices. The quark condensate is denoted by $\big<\bar{\psi}_q \psi_q \big>$, which is known as an
order parameter of chiral symmetry breaking. The dressed quark propagator for different flavors of
$q$  is defined by
\begin{eqnarray}
    \label{eq:gap1}
    S^{-1}_q (k) &=& k\!\!\!/ - M_q + i \epsilon,
\end{eqnarray}
where, $\epsilon > 0$, and infinitesimal as usual.
The formation of the (dressed quark)-(dressed antiquark) bound $\rho$ meson state can be obtained by solving the Bethe-Salpeter equation (BSE) in the random phase approximation, which is
equivalent to the ladder approximation. The solution to BSE in the vector channel is
given by a two-body $t$-matrix of the interaction channel. The summation of the bubble diagram for
the pion and $\rho$ meson from infinite interaction can be written as
\begin{eqnarray}
    \label{eq:vacuumNJL3}
    t_\pi (p^2) &=& \frac{-2i G_\pi}{1 + 2 G_\pi \Pi_{\pi} (p^2)},~~~~~~
    t_{\rho (\omega)}^{\mu \nu} (p^2) = \gamma^{\mu}  \Bigg( \frac{-2iG_{\rho (\omega)}}{1 + 2 G_{\rho (\omega)} \Pi_{\rho (\omega)} (p^2) }\Bigg) \gamma^\nu  \Bigg[ g^{\mu \nu} + 2 G_{\rho(\omega)} \Pi_{\rho(\omega)} (p^{2}) \frac{p^{\mu} p^{\nu}}{p^2}\Bigg].
\end{eqnarray}
The polarization insertions (bubble diagrams) for the pion and vector mesons are given by 
\begin{eqnarray}
    \label{eq:vacuumNJL4}
    \Pi_\pi (p^2) \delta_{ab} &=& 3i \int \frac{d^4k}{(2\pi)^4} \mathrm{Tr} \Big[ \gamma_5 \tau_a
S_q (p+k)
    \gamma_5 \tau_b S_q (k) \Big], \nonumber \\
      \Pi_{\rho (\omega)} (p^2) P^{\mu \nu} \delta_{ab} &=& 3i \int \frac{d^4k}{(2\pi)^4}
\mathrm{Tr} \Big[
      \gamma^\mu \tau_a S_q (p+k) \gamma^\nu \tau_b S_q (k) \Big],
\end{eqnarray}
respectively, where $P^{\mu \nu} = \Big( g^{\mu \nu}-\frac{p^\mu p^\nu}{p^2}\Big)$, and the trace
operates over the Dirac and flavor indices. Note that the $\rho$ meson mass is defined by
the pole position
in the corresponding two-body $t$-matrix.

\subsection{$\rho$ meson masses}
The pion, $\rho$, and $\omega$ meson masses can be straightforwardly determined from the pole positions of
the corresponding $t$-matrix in Eq.~(\ref{eq:vacuumNJL3}), and one has
\begin{eqnarray}
    \label{eq:vacuumNJL5}
     1 + 2 G_\pi \Pi_\pi (p^2 = m_\pi^2) &=& 0, \\
    1 + 2 G_\rho \Pi_\rho (p^2 = m_\rho^2) &=& 0, \\
    \label{eq:vacuumNJL5a}
    1 + 2 G_\omega \Pi_{\omega} (p^2 = m_{\omega}^2) &=& 0,
\end{eqnarray}
where they are determined by the bound state pole positions of the $t$-matrix. The $m_\pi$, $m_\rho$, and $m_\omega$ in Eqs.~(\ref{eq:vacuumNJL5})-(\ref{eq:vacuumNJL5a}) denote the masses of the pion, rho, and omega mesons,
respectively.

To determine the meson-quark coupling constant, we expand the $t$-matrix in
Eq.~(\ref{eq:vacuumNJL3}) around the
pole $p^2 = m_{\rho (\omega)}^2$, and we can obtain the meson wave function
renormalization constants or the meson-quark coupling constants by using
\begin{eqnarray}
    \label{eq:vacuumNJL8}
     Z_\pi^{-1} = \Big[ g_{\pi q q}\Big]^{-2} &=& - \frac{\partial \Pi_{\pi} (p^2)}{\partial p^2}
\Bigg|_{p^2 = m_{\pi}^2}, \\
   Z_{\rho (\omega)}^{-1} = \Big[ g_{\rho (\omega) q q}\Big]^{-2} &=& - \frac{\partial \Pi_{\rho (\omega)}
(p^2)}
{\partial p^2} \Bigg|_{p^2 = m_{\rho (\omega)}^2},
\end{eqnarray}
where $g_{\pi qq}$ and $g_{\rho (\omega) qq}$  represent the corresponding meson-quark coupling
constants.
Next, we extend these free-space NJL expressions
to the nuclear medium in Sec.~\ref{sec:vectormediumnNJL}. Before explaining the properties of the
$\rho$ meson in the nuclear medium, in Sec.~\ref{sec:NMNJL}, we explain the nuclear matter
description in the NJL model.

\section{NJL model nuclear matter} \label{sec:NMNJL}
In this section, we present the extension of the SU(2) flavor NJL model Lagrangian to symmetric nuclear matter, applying the Fierz transformation~\cite{Ishii:1993rt}, where the
four-fermion interactions can be decomposed as a chiral symmetric linear combination of the form
$\sum_i G_i \big( \bar{\psi}_q \Gamma_i \psi_q \big)^2$. Note that the NJL nuclear matter model is
characterized by the nuclear (baryon) density $\rho_B$. After performing the quark
bilinear\,\footnote{The quark bilinear $\big(\bar{\psi}_q \psi_q \big)$ and $\big( \bar{\psi}_q
\gamma^\mu \psi_q \big)$ can be decomposed as $\big<\rho_B\mid \bar{\psi}_q \psi_q \mid \rho_B \big>
+ \big(:\bar{\psi}_q \psi_q:\big)$  and $\big<\rho_B\mid \bar{\psi}_q \gamma^\mu \psi_q \mid \rho_B
\big> + \big(  : \bar{\psi}_q \gamma^\mu \psi_q :\big)$. For details, see Ref.~\cite{Bentz:2001vc}.}
into the effective NJL Lagrangian in Eq.~(\ref{eq:vacNJL1}), one
has~\cite{Bentz:2001vc,Gifari:2024ssz}
\begin{eqnarray} 
\label{eq:NJL7}
\mathscr{L}_{\mathrm{NM-NJL}} &=& \bar{\psi}_q \big( i \partial\!\!\!/ - M_q - V\!\!\!\!/
\big)\psi_q -
\frac{\big( M_q-m_q\big)^2}{4G_\pi} + \frac{V_\mu V^\mu}{2G_\omega} + \mathscr{L}_{I},
\end{eqnarray}  
where $\mathscr{L}_{I}$ stands for the interaction Lagrangian. Through the Fierz
transformation and charge conjugation,
$\mathscr{L}
_{I}$ in Eq.~(\ref{eq:NJL7}) is decomposed into a sum of the $qq$ interactions of the
isoscalar-scalar ($0^+, T=0$) and isovector-axial vector ($1^+, T=1$), so that $\mathscr{L}_{I}$ can
be written as
\begin{eqnarray}
\mathscr{L}_{I,qq} &=& G_s \big[ \bar{\psi}_q \gamma_5 C \tau_2 \beta_A \bar{\psi}_q^T \big] \big[
\psi^T C^{-1} \gamma_5 \tau_2 \beta_A \psi_q \big] + G_a \big[ \bar{\psi}_q \gamma_\mu C \tau_i
\tau_2 \beta_A \bar{\psi}_q^T \big] \big[ \psi_q^T C^{-1} \gamma^\mu \tau_2 \tau_i \beta_A \psi_q
\big],
\end{eqnarray}
where $\beta_A = \sqrt{\frac{3}{2}} \lambda_A$ with $A = 2, 5, 7$,
$C = i \gamma_2 \gamma_0$, and
$G_s$ and $G_a$ are the coupling constants of the scalar and axial vector $qq$ interactions,
respectively. It is worth noting that the scalar coupling constant \( G_s \) is tuned to reproduce
the free nucleon mass. The isoscalar-vector mean field and the dynamical effective quark mass can
be, respectively, introduced by
\begin{eqnarray}
\label{eq:NJL8}
V^\mu &=& 2 G_\omega \big< \rho_B \mid \bar{\psi}_q \gamma^\mu \psi_q \mid \rho_B \big> = 2
\delta^{0\mu}
G_\omega \big< \psi_q^\dagger \psi_q \big>, \\
\label{eq:NJL8b}
M_q &=& m_q - 2 G_\pi \big< \rho_B \mid \bar{\psi}_q \psi_q \mid \rho_B \big>,
\end{eqnarray}  
where the vector potential can be defined as $V=(V_0,\mathbf{0})$ in symmetric nuclear
matter at rest.
Then, by the stable (minimum) condition $\partial \mathcal{E} / \partial V_0 =$ 0,
the value of $V_0$ is given as
   $ V_0 = 6 G_\omega \rho_B $,
where the $\rho_B = 2 p_F^3 /3 \pi^2$ stands for the baryon number density (nuclear matter density).
Notably, the dynamical quark mass for fixed baryon density must satisfy
the stable (minimum) condition $\partial
\mathscr{E} /\partial M_q = \frac{\big(M_q -m_q\big)}{2G_\pi} + \big<\rho_B \mid \bar{\psi}_q \psi_q
\mid \rho_B \big> = 0$ to have a similar in-medium dynamical quark mass as in Eq.~(\ref{eq:NJL8}).

Using the standard hadronization technique~\cite{Bentz:2001vc}, the effective potential for
symmetric nuclear
matter can be determined from the effective NJL Lagrangian. In the mean-field approach (MFA), the
energy density of the symmetric nuclear matter is introduced by
\begin{eqnarray}
\label{eq:NJL9} 
\mathscr{E} = \mathscr{E}_V- \frac{V_0^2}{4G_\omega} + 4 \int \frac{d^3p}{(2\pi)^3} \Theta
\big(p_F - |\mathbf{p}|\big)\, \epsilon_N,
\end{eqnarray}  
where $\epsilon_N = \sqrt{M_N^{*2} + \mathbf{p}^2} + 3V_0 \equiv E_N + 3V_0$
with $M_N^* (M_q \equiv M_q^*)$ is the in-medium nucleon mass obtained from the pole of the
quark-diquark $t$-matrix and the nucleon Fermi momentum $p_F$, which is given by
$p_F^2 = \big[\big(\mu_N -3V_0\big)^2 - M_N^{* 2} \big]$, with $\mu_N$ being
the chemical potential
of nucleon. The vacuum contribution of the quark is given by
\begin{eqnarray}
\label{eq:NJL10}    
\mathscr{E}_V &=& 12i \int \frac{d^4k}{(2\pi)^4} \ln \Bigg( \frac{k^2 -M_q^2 + i\epsilon}{k^2 -
M_0^2 + i \epsilon}\Bigg) + \frac{\left( M_q-m_q\right)^2}{4G_\pi} - \frac{\left(
M_0-m_q\right)^2}{4G_\pi},
\end{eqnarray}  
where $M_0$ is the vacuum constituent quark mass at zero density. Using the expression of the energy
density for the NJL symmetric nuclear matter, the (negative of) energy per
nucleon for symmetric nuclear matter with free space nucleon mass $M_{N0}$ can be given by
\begin{eqnarray}
\frac{E_B}{A} &=& \frac{\mathscr{E}}{\rho_B} - M_{N0}.
\end{eqnarray}  
As mentioned before, the stability of symmetric
nuclear matter of the NJL model is guaranteed by reproducing the energy per nucleon $E_B/A
=-15.7$ MeV at saturation density $\rho_0 = 0.16$ fm$^{-3}$. Now we study the $\rho$ meson
properties in the nuclear medium by using the quark properties in symmetric nuclear
matter calculated by the NJL model as inputs.

\begin{table*}[t]
	\begin{ruledtabular}
		\renewcommand{\arraystretch}{1.3}
		\caption{Density dependence of constituent quark mass,
		$\rho$ meson mass,
$\rho$ decay constant, $\rho$-quark coupling constant, $\rho^+$ magnetic moment, $\rho^+$ quadrupole
moment, $\rho^+$ charge radius, quark condensate, and nucleon effective mass calculated in the NJL
model. All units are in MeV.
The units of $Q_\rho^*$ and $\big<r_\rho^* \big>$ are in fm$^2$ and fm,
respectively.}
		\label{tab:NJL2}
		\begin{tabular}{cccccccccc}
		  $\rho_B/\rho_0$  & $M_q^*$ & $m_{\rm \rho}^*$ &$g_{\rm \rho qq}^{*}$ & $\mu_\rho^* (\mu_N)$ & $Q_\rho^*$ &$\big< r_\rho^*\big>$ &  $-\langle \bar{u}u \rangle^{*1/3}$ & $M_N^*$\\ 
\hline  
		 $0.00$ &  400  & 770 & 2.638  & 2.53  & -0.057 & 0.67 & 171 & 934.36 \\
		 $0.50$ &  360 & 726  & 2.537   & 2.41  & -0.065 & 0.71 & 165 & 842.54\\
		 $1.00$ &  328 & 694   & 2.459   & 2.33  & -0.071& 0.75 & 160 & 778.03\\
		 $1.50$ &  306 & 673  &  2.410  & 2.28  & -0.076 & 0.77 & 156 & 738.98\\
          $2.00$ &  289 & 658  & 2.373   & 2.26 & -0.080 & 0.79 &  153 & 715.34\\
		\end{tabular}
		\renewcommand{\arraystretch}{1}
	\end{ruledtabular}
\end{table*}

\section{$\rho$ meson in nuclear medium} \label{sec:vectormediumnNJL}
Here, we present the formulation of the in-medium properties of the $\rho$ meson.
The NJL dynamical quark mass in symmetric nuclear matter is expressed as
\begin{eqnarray}
    \label{eq:NJL1}
    M_q^* &=& m_q + 4i G_\pi \mathrm{Tr} \big[ S_q^* (p^*) \big],
\end{eqnarray}
where $M_q^{*}$ represents the dynamical quark effective masses, which corresponds to $M_q$ in
Eq.~(\ref{eq:NJL8b}). The
coupling constant \( G_\pi \) represents the scalar interaction strength and is taken to be the same
as in free space. Note that here the $\Lambda_{\mathrm{IR}}$ is set to zero in the deconfined
phase, when we consider a nuclear (quark) matter. This implies that nucleon decay into quarks may occur at
an unphysical threshold. The in-medium dressed light quark propagators in Eq.~(\ref{eq:NJL1}) can
be expressed by
\begin{eqnarray}
\label{Eq:NJL2}
S_q^{*} (k^{*}) &=&
\frac{\big(k\!\!\!/^{*} + M^{*}_q\big)}{\big(k^{*2} -M_q^{*2} + i \epsilon\big)},
\end{eqnarray}  
where the asterisk symbol in Eq.~(\ref{Eq:NJL2}) stands for the in-medium quantity which
enters the light quark momentum $k^\mu$ by $k^{*\mu} = k^\mu + V^\mu$, due to the vector meson mean
field $V^\mu =\left( V^0 = V_0,\mathbf{0} \right)$. Note that the medium modifications of the
space
component of the light quark momentum $k^{*\mu}$ can be ignored, since the contribution is
insignificant~\cite{Krein:1998vc}. It is to be noted that, after substituting the in-medium modifications of the quark
propagator into the constituent quark mass formula in Eq.~(\ref{eq:NJL1}),
we can shift the variable of the integral to eliminate the vector potential that enters the light quark momentum.

We now present the description of the bound state of the mesons in the nuclear medium. Shortly, the
in-medium reduced $t$-matrices
for the $\pi$ and $\rho (\omega)$ mesons are respectively obtained by
\begin{eqnarray}
  \label{eq:NJL3}
     t_\pi^{*} (p^*) &=& \frac{-2i G_\pi}{1+ 2 G_\pi \Pi_\pi^{*} (p^{*2})},~~~~~
     t_{\rho (\omega)}^{*\mu \nu} (p^{*}) = \frac{-2iG_\rho}{1 + 2 G_\rho \Pi_V^{*} (p^{*2})} \Bigg[
g^{\mu \nu}
     + 2 G_{\rho(\omega)} \Pi^{*}_{\rho (\omega)} (p^{*2}) \frac{p^{*\mu} p^{*\nu}}{p^2}\Bigg],
\end{eqnarray}  
where $G_\rho$ is the four-fermion coupling constant for the vector meson channels and the
polarization
insertions, which are the so-called bubble diagrams in the nuclear medium, can be introduced by
\begin{eqnarray} 
    \label{eq:NJL4} 
    \Pi^{*}_\pi (p^{*2}) &=& 6i \int \frac{d^4k}{(2\pi)^4} \mathrm{Tr} \Big[ \gamma_5 S_q^* (k^*)
\gamma_5 S_q^{*} (k^{*}+p^{*}) \Big],\nonumber \\
    \Pi^{*}_{\rho(\omega)} (p^{*2}) P_T^{\mu \nu} &=& 6i \int \frac{d^4k}{(2\pi)^4} \mathrm{Tr} \Big[ \gamma^\mu
S_q^* (k^*) \gamma^\nu S_q^* (k^* + p^*)\Big],
\end{eqnarray}    
where $P_T^{\mu \nu} = \big(g^{\mu \nu} - q^\mu q^\nu/q^2 \big)$, and $\mathrm{Tr}$ only operates to
the Dirac
matrices. After extracting the polarization insertion for the vector mesons, we can
straightforwardly determine the in-medium meson masses for the $\pi$ and $\rho (\omega)$ mesons
as the pole positions of the $t$-matrix, which are given by
\begin{eqnarray}
    \label{eq:NJL5}
    1 + 2 G_\pi \Pi^{*}_\pi (p^{*2} = m_\pi^{*2}) &=& 0,\\
    1 + 2 G_{\rho(\omega)} \Pi^{*}_{\rho (\omega)} (p^{*2} = m_{\rho(\omega)}^{*2}) &=& 0,
\end{eqnarray}
where $m_\pi^*$, and $m_{\rho (\omega)}^*$ are the pion and vector meson $\rho (\omega)$ effective
masses, respectively. Next, the  pion-quark and $\rho (\omega)$-quark coupling constants
in the nuclear medium can be computed by taking the first derivative of the
polarization insertion with respect to $p^{*2}$, which yields
\begin{eqnarray}
    \label{eq:NJL6} 
\big[Z_\pi^*\big]^{-1} =\big[ g_{\pi q q}^{*} \big]^{-2} &=& - \frac{\partial \Pi_\pi^{*}
(p^{*2})}{\partial
p^{*2}} \Bigg|_{p^{*2} = m_\pi^{*2}},\\
\label{eq:NJL6b}
\big[ Z_{\rho (\omega)}^*\big]^{-1} = \big[ g_{\rho(\omega) q q}^{*} \big]^{-2} &=& - \frac{\partial
\Pi_{\rho
(\omega)}^{*} (p^{*2})}{\partial p^{*2}} \Bigg|_{p^{*2} = m_{\rho(\omega)}^{*2}}.
\end{eqnarray}  
In the next section, we discuss the formulation of the $\rho$ meson EMFFs in the nuclear
medium. Furthermore, we explain the in-medium modifications of the charge radius, the magnetic
moment, and the quadrupole moment of the $\rho$ meson, which represent the main goal of this work.

\section{In-medium $\rho$ meson EMFFs}
\label{sec:ffv}
Here we describe the $\rho$ meson EMFFs in the nuclear medium in the covariant NJL model. Shortly, the generic expression for the electromagnetic current for the $\rho$ meson can be parametrized in
terms of three form factors,
\begin{eqnarray}
    \label{eq:vff1}
   \mathcal{J}_\rho^{\mu, \alpha \beta} (p'^{*},p^{*}) &=& \Big[ g^{\alpha \beta} F_{1}^{*\rho} (Q^2)
- \frac{q^{\alpha *} q^{\beta *}}{2 m_\rho^{* 2}} F_{2}^{*\rho} (Q^2) \Big] \left( p'^{*} +
p^{*}\right)^\mu - \Big[ q^{\alpha *} g^{\mu \beta} - q^{\beta *} g^{\mu \alpha}\Big] F_{3}^{*\rho}
(Q^2),
\end{eqnarray}
where the superscript of the Lorentz indices $\alpha$ and $\beta$ respectively correspond to
the polarizations of the incoming and outgoing $\rho$ meson, while $\mu$ corresponds to
the photon polarization. $F_{1}^{*\rho}
(Q^2)$, $F_{2}^{*\rho} (Q^2)$, and $F_{3}^{*\rho}(Q^2)$ are the in-medium $\rho$ meson EMFFs
that we want to extract using the NJL model.

These form factors can be written in terms of three Sachs form factors
for the $\rho$ meson in the nuclear medium: the
charge/electric form factor [$G_C^* (Q^2)$], which is related to the charge distribution, magnetic
form factor [$G_M^* (Q^2)$], which is associated with the magnetization distribution, and quadrupole
form factor [$G_Q^* (Q^2)$], which reflects the shape or spatial symmetry of the $\rho$ mesons.
Explicitly, they are defined by
\begin{eqnarray}
    \label{eqvff2}
    G_C^{*} (Q^2) &=& F_{1}^{*\rho} (Q^2) + \frac{2}{3} \eta G_Q^* (Q^2), \\
     G_Q^{*}(Q^2) &=& F_{1}^{*\rho} (Q^2) + \big(1 + \eta \big) F_{2}^{*\rho} (Q^2) - F_{3}^{*\rho} (Q^2), \\
    G_M^{*} (Q^2) &=& F_{3}^{*\rho} (Q^2),
\end{eqnarray}
where $\eta = Q^{2} / 4 m_{\rho}^{* 2}$. It is worth noting that all form factors above
are dimensionless.

\begin{figure}[t]
\centering
\includegraphics[width=0.85\columnwidth]{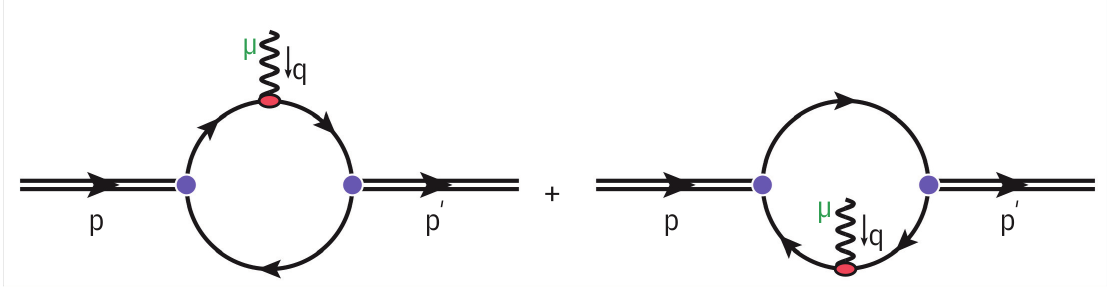} 
\caption{ \label{fig6a} Diagrammatic representations of the electromagnetic interaction with the
$\rho$ meson.
In each panel, the BSE vertices are represented by the filled pair circles (purple), while the
quark-photon vertex is represented by a filled oval (red), where we can assume that in the left
figure, the photon hits the quark and, in the right figure, the photon hits the antiquark.}
\end{figure}

In the NJL model, based on the diagrams in Fig.~\ref{fig6a}, the electromagnetic current for the
$\rho$ meson in the nuclear medium can be expressed by
\begin{eqnarray}
    \label{eqvff2a}
    \mathcal{J}_{\rho,ij}^{\mu, \alpha \beta} (p',p) &=& i \int \frac{d^4k}{(2\pi)^4} \mathrm{Tr}
\Big[
    \bar{\Gamma}_\rho^{\beta, j} S^*_q(p'^*+k^*) \Lambda^{\mu *}_{\gamma Q} (p',p) S^*_q(p^*+k^*)
\Gamma_{\rho}^{\alpha, i} S^*_q(k^*) \Big] \nonumber \\
    &+& i \int \frac{d^4k}{(2\pi)^4} \mathrm{Tr} \Big[ \Gamma_\rho^{\alpha, i} S^*_q(k^*-p^*)
\Lambda^{\mu *}
    _{\gamma Q} (p',p)  S^*_q(k^*-p^*) \bar{\Gamma}_\rho^{\beta, j} S^*_q(k^*) \Big], \nonumber \\
     &=& 2i \int \frac{d^4k}{(2\pi)^4} \mathrm{Tr} \Big[ \bar{\Gamma}_\rho^{\beta, j}
S^*_q(p'^*+k^*)
     \Lambda^{\mu *}_{\gamma Q} (p',p) S^*_q(p^*+k^*) \Gamma_{\rho}^{\alpha, i} S_q^{*T} (-k^*)
\Big],
\end{eqnarray}
where the trace, $\mathrm{Tr}$, operates over the Dirac, color, and isospin indices, the
superscript $T$ stands for
the transpose in the Dirac space, and $\Lambda^{\mu *}_{\gamma Q} (p',p)$ stands for the
in-medium dressed quark-photon
vertex, and is defined as
\begin{eqnarray}
    \label{eqvff2b}
    \Lambda_i^{\big(\mathrm{BSE}\big) \mu *} (Q^2) &=& \gamma^\mu F^*_{1i} (Q^2) +
\frac{i\sigma^{\mu \nu}
    q_\nu}{2M} F^*_{2i} (Q^2),
\end{eqnarray}
where the subscript $i = \big(\rho, \omega \big)$ is the dressed quark form factor obtained from
the inhomogeneous BSE. Note that $F^*_{1\omega} (Q^2) = F^*_{1\omega} (Q^2) =$ 1 and $F^*_{2\rho}
(Q^2) = F^*_{2\omega} (Q^2) =$ 0, for the pointlike quark. This implies $F_{2Q}^* (Q^2) =
\frac{1}{6} F_{2\omega}^* (Q^2) \pm \frac{1}{2} F_{2 \rho}^* (Q^2)= 0$ where $Q = (U,D)$ (dressed
quarks). However,
$f_1^{T *} (Q^2), f_2^{T *} (Q^2)$, and $f_3^{T *} (Q^2)$ are not zero as can be seen in
Refs.~\cite{Carrillo-Serrano:2015uca,Cloet:2014rja}. It is worth noting that $f_3^T (Q^2) = 0$ as it
preserves charge conservation.

After full derivations and some calculations of the in-medium EM current
for the $\rho$ meson in the NJL model, Eq.~(\ref{eqvff2a}),
and comparing it with the current in
Eq.~(\ref{eq:vff1}), we can extract the $\rho$ meson EMFFs in the nuclear medium.
The final expressions for the $\rho$ meson EMFFs in the NJL model are respectively
given by
\begin{eqnarray}
    \label{eq:vff3}
    F_{1}^{*\rho} (Q^2) &=& \left[ F_{1U}^* (Q^2) -F_{1D}^* (Q^2) \right] f_1^{V *} (Q^2) + \left[
F_{2U}^*
    (Q^2) - F_{2D}^* (Q^2) \right] f_{1}^{T *} (Q^2), \\
     \label{eq:vff3a}
     F_{2}^{*\rho} (Q^2) &=& \left[ F_{1U}^* (Q^2) -F_{1D}^* (Q^2) \right] f_2^{V *} (Q^2)  +
\left[ F_{2U}^*
     (Q^2) - F_{2D}^* (Q^2) \right] f_{2}^{T *} (Q^2),\\
      \label{eq:vff3b}
      F_{3}^{*\rho} (Q^2) &=& \left[ F_{1U}^* (Q^2) -F_{1D}^* (Q^2) \right] f_3^{V *} (Q^2)  +
\left[ F_{2U}^*
      (Q^2) - F_{2D}^* (Q^2) \right] f_{3}^{T *} (Q^2),
\end{eqnarray}
where the subscripts $1,2$, and 3 stand for the three form factors given in Eq.~(\ref{eq:vff1})
and $f_i^{V*}$ is the
body form factor for vector part with $i=1,2,3$. The vector body form factors in the Schwinger
proper time regularization scheme are given by
\begin{eqnarray}
    \label{eq:bodyff}
    f_1^{V*} (Q^2) &=& - \frac{3g_{\rho qq}^{*2}}{16 \pi^2}
\int_{\tau_{\mathrm{UV}}}^{\tau_{\mathrm{IR}}} d\tau
    \int_0^1 dx \int_{-x}^{x} dy \Big( \frac{4}{\tau} \big(1-x\big) - 2x m_\rho^{*2} \Big) \exp\Big[
-\tau \big( (x^2-x) m_\rho^{*2} + \frac{1}{4} (x^2 -y^2) Q^2 + M_q^{*2} \big) \Big]  \nonumber \\
    &+& \frac{3 g_{\rho qq}^{*2}}{4\pi^2} \int_{\tau_{\mathrm{UV}}}^{\tau_{\mathrm{IR}}} d\tau
\int_0^1 dx
    \frac{\exp \Big[ -\tau((x-x^2)Q^2 + M_q^{*2} \Big]}{\tau}, \\
    f_2^{V*} (Q^2) &=& \frac{3 g_{\rho qq}^{*2} m_\rho^{*2}}{4\pi^2} \int_{\tau_{\mathrm{UV}}}
    ^{\tau_{\mathrm{IR}}} d\tau \int_0^1 dx \int_{-x}^x dy \big( x^2 -y^2 \big) \big( 1-x \big)
\exp\Big[ -\tau \big( (x^2-x) m_\rho^{*2} + \frac{1}{4} (x^2 -y^2) Q^2 + M_q^{*2} \big) \Big], \\
    f_3^{*V} (Q^2) &=& -\frac{3 g_{\rho qq}^{*2}}{16 \pi^2}
\int^{\tau_{\mathrm{IR}}}_{\tau_{\mathrm{UV}}} d\tau
    \int_0^1 dx \int_{-x}^{x} dy \Big[ \frac{4}{\tau} (x+1)-2x (1+2x) m_\rho^{*2} - (x^2 -y^2) Q^2
\Big] \nonumber \\
    &\times& \exp\Big[ -\tau \big( (x^2-x)m_\rho^{*2} + \frac{1}{4} (x^2 -y^2)Q^2 + M_q^{*2} \big)
\Big]
    \nonumber \\
    &+& \frac{9 g_{\rho qq}^{*2}}{4 \pi^2}  \int_0^1 dx
\int_{\tau_{\mathrm{UV}}}^{\tau_{\mathrm{IR}}}
    \frac{d\tau}{\tau} \exp\Big[ -\tau \big( (x-x^2)Q^2 + M_q^{*2} \big)\Big],
\end{eqnarray}
while the tensor body form factors are given by
\begin{eqnarray}
    \label{eq:bodyffb}
    f_{1}^{T*} (Q^2) &=& - \frac{3 g_{\rho q q}^{*2} Q^2}{16 \pi^2} \int_0^1 dx \int_{-x}^{x} dy
    \int_{\tau_{\mathrm{UV}}}^{\tau_{\mathrm{IR}}} d\tau \exp \Big[ -\tau \big( (x^2-x) m_\rho^{*2}
+ \frac{1}{4} (x^2 -y^2) Q^2 + M_q^{* 2} \big)\Big],\\
    f_2^{T*} (Q^2) &=& \frac{3g_{\rho qq}^{*2}m_\rho^{*2}}{4\pi^2} \int_0^1 dx \int_{-x}^{x} dy
    \int_{\tau_{\mathrm{UV}}}^{\tau_{\mathrm{IR}}} d\tau \big(1-x\big) \exp \Big[ -\tau \big(
(x^2-x) m_\rho^{*2} + \frac{1}{4} (x^2-y^2) Q^2 + M_q^{*2} \big) \Big], \\
    f_3^{T*} (Q^2) &=& - \frac{3 g_{\rho qq}^{*2}}{16 \pi^2} \int_0^1 dx \int_{-x}^{x} dy
    \int_{\tau_{\mathrm{UV}}}^{\tau_{\mathrm{IR}}} d\tau \Big[ xQ^2 - 2m_\rho^{*2} \Big] \exp \Big[
-\tau \big( (x^2-x) m_\rho^{*2} + \frac{1}{4} (x^2 -y^2) Q^2 + M_q^{*2} \big)\Big] \nonumber \\
    &+& \frac{3g_{\rho qq}^{*2}}{4\pi^2} \int_0^1 dx \int_{\tau_{\mathrm{UV}}}^{\tau_{\mathrm{IR}}}
\frac{d\tau}
    {\tau} \exp \Big[ -\tau \big( (x-x^2)Q^2 + M_q^{*2} \big) \Big].
\end{eqnarray}

The inhomogeneous BSE-dressed quark form factors in the nuclear medium can be defined by
\begin{eqnarray}
\label{eq:vff4}
F_{1U}^* (Q^2) &=& \frac{1}{6} F_{1\omega}^* (Q^2) + \frac{1}{2} F_{1 \rho}^* (Q^2),  \\
F_{1D}^* (Q^2) &=& \frac{1}{6} F_{1\omega}^* (Q^2) - \frac{1}{2} F_{1 \rho}^* (Q^2), \\
F_{2U}^* (Q^2) &=& \frac{1}{6} F_{2\omega}^* (Q^2) + \frac{1}{2} F_{2 \rho}^* (Q^2),  \\
F_{2D}^* (Q^2) &=& \frac{1}{6} F_{2\omega}^* (Q^2) - \frac{1}{2} F_{2 \rho}^* (Q^2),
\end{eqnarray}
where the dressed quark form factors are given by
   $ F_{1 \rho (1\omega)}^* (Q^2) = 1/\big[1 + 2 G_{\rho (\omega)} \Pi_{\rho (\omega)}^* (p^{* 2}) \big]$, 
    $F_{2 \rho (2\omega)}^* (Q^2) = 0 $,
and $F_{1\rho}^* (Q^2)$ and $F_{1\omega}^* (Q^2)$ have poles at $ p^{* 2} = m_\rho^{* 2}$ and
$p^{* 2} = m_\omega^{* 2}$, respectively. It is worth noting that the BSE kernel of the NJL model
does not generate the Pauli form factors $F_{2\rho(2\omega)}^* (Q^2)$ because of the absence of the tensor-tensor four-fermion interaction in the present approach.

Through the $\rho$ meson EMFFs in nuclear medium, we can straightforwardly
compute the static electromagnetic quantities of the $\rho$ meson in the nuclear medium, such as
the magnetic moment $\mu_\rho^*$, quadrupole moment $\mathcal{Q}_\rho^*$, and the squared
charge radius $\big< r_C^{*2} \big>$. These static quantities can be determined, respectively, by
\begin{eqnarray}
    \mu^*_\rho &=& G^*_M (Q^2 =0) \frac{M^*_N}{m^*_\rho}, \\
    \mathcal{Q}^*_\rho &=& \frac{G^*_Q (Q^2=0)}{m^{* 2}_\rho}, \\
    \label{eqrad}
    \big< r_C^{*2} \big> &=& -6 \frac{\partial G^*_C (Q^2)}{\partial Q^2} \Bigg|_{Q^2 =0},
\end{eqnarray}
where $M_N^*$ stands for the effective nucleon mass. The values for both $M_N^*$ and $m_\rho^*$
masses for various matter densities are given in
Table~\ref{tab:NJL2}. It is worth noting that the charge is normalized by
$G_C^* (Q^2=0) =  1.$

\section{Numerical Result} \label{sec:MR}
Here, we present the numerical results for the $\rho$ meson properties and their EMFFs, such as
the charge radius $\big< r_\rho^* \big> \equiv \big<r^{*2}_C \big>^{1/2}$ (See Eq.~(\ref{eqrad})), the charge form factor $G_C^*(Q^2)$, magnetic form factor
$G_M^* (Q^2)$, and quadrupole moment form factor $G_Q^* (Q^2)$)
in the nuclear medium. The NJL model parameter
used in the  present work is $G_\pi$, $G_\omega$, $G_\rho$, $G_a$, and $G_s$, as used in
Refs.~\cite{Hutauruk:2021kej,Hutauruk:2016sug,Bentz:2001vc}. In this work, the regularization
parameter $\Lambda_{\mathrm{IR}}$ in the Schwinger proper-time regularization scheme  is fixed at
240 MeV, which is comparable to $\Lambda_{\mathrm{QCD}}$,
while the free space (vacuum) constituent quark mass is taken as $M_0 = 400$ MeV for $u$ and $d$ quarks.It is worth mentioning that the fixing of the dynamical quark mass in the NJL model is somewhat arbitrary; however, the results obtained for other value choices of $M_0$ are qualitatively expected to be quite similar, as indicated in Ref.~\cite{Bentz:2001vc}.

The remaining free space parameters are fitted to the physical pion mass $m_\pi =$ 140 MeV,
physical kaon mass $m_K =$ 495
MeV, nucleon mass $M_N = M_{N0} =$ 940 MeV, and $\rho$ meson mass $m_\rho =$ 770 MeV, together with
the pion weak-decay constant $f_\pi =$ 93 MeV. The fitted results give an ultraviolet cutoff
$\Lambda_{\mathrm{UV}} =$ 645 MeV, pion coupling constant $G_\pi =$ 19.04 $\times $ $10^{-6}$
MeV$^{-2}$, axial-vector diquark coupling constant $G_a =$ 2.8 $\times$ $10^{-6}$ MeV$^{-2}$, scalar
diquark coupling constant $G_s =$ 7.49 $\times$ $10^{-6}$ MeV$^{-2}$, and strange constituent quark
mass $M_s =$ 611 MeV. The $\phi$ meson mass is taken $m_\phi =$ 1001 MeV, scalar diquark mass
$M_{sdi} =$ 687 MeV, and axial diquark mass $M_{adi} =$ 1027 MeV. It is worth noting that the coupling
constants $G_s$ and $G_a$ are determined by solving the Faddeev equation to reproduce the nucleon
mass in vacuum, $M_N$, and axial coupling constant of nucleon $g_A = 1.267$, as employed in
Ref.~\cite{Bentz:2001vc}.

The parameters for symmetric nuclear matter are determined by fitting the (negative of) energy per nucleon, $E_B/A = -15.7$ MeV, at the saturation baryon density $\rho_0 = 0.16$
fm$^{-3}$. This yields a value of
$G_\omega = 6.03 \times 10^{-6}$ MeV$^{-2}$. The current quark mass value (in vacuum)
for the up quark (and down quark) is taken as $m_u = m_d = 16$ MeV (SU(2) isospin symmetry), while the current mass of the
strange quark is set to $m_s = 356$ MeV, following
Refs.~\cite{Hutauruk:2021kej,Hutauruk:2016sug,Bentz:2001vc}.

\subsection{In-medium $\rho$ meson properties}
\begin{figure*}[ht]
\centering
\includegraphics[width=0.49\columnwidth]{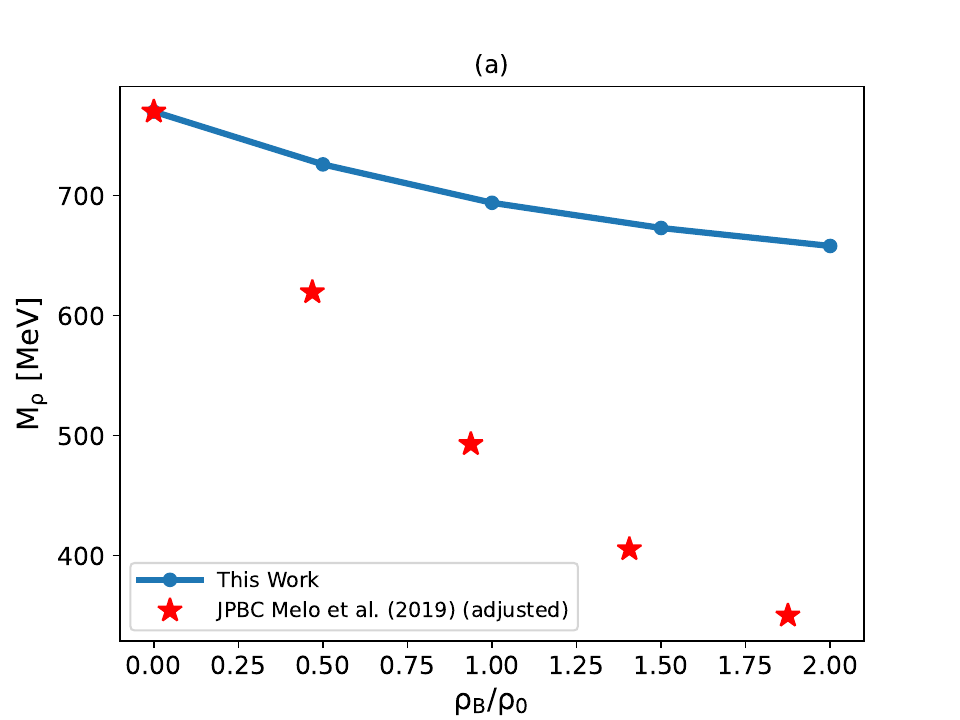} 
\includegraphics[width=0.49\columnwidth]{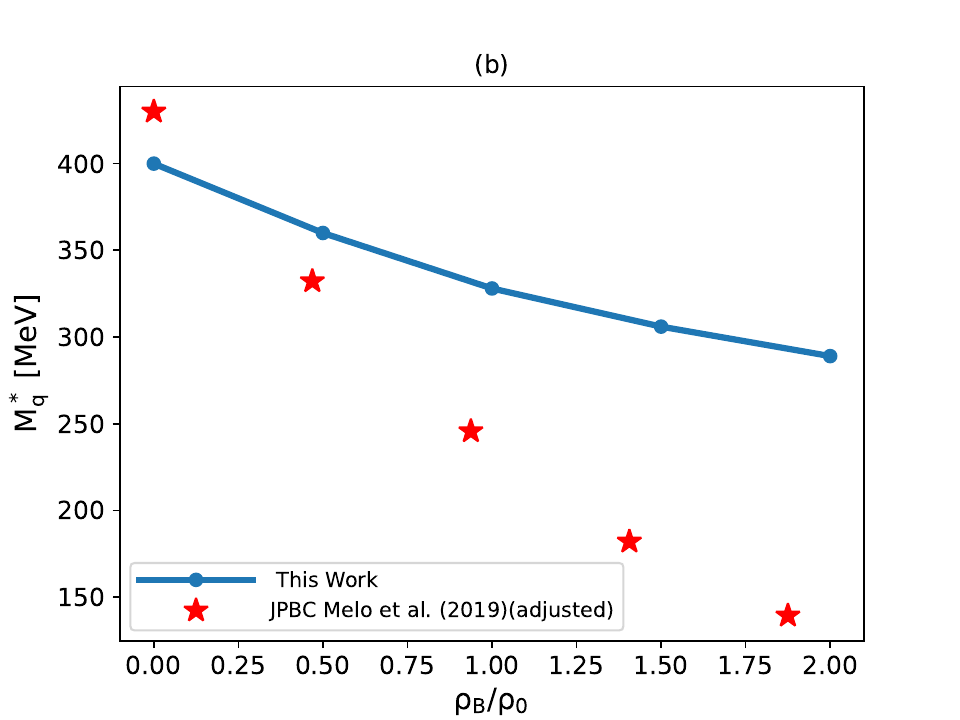} 
\includegraphics[width=0.49\columnwidth]{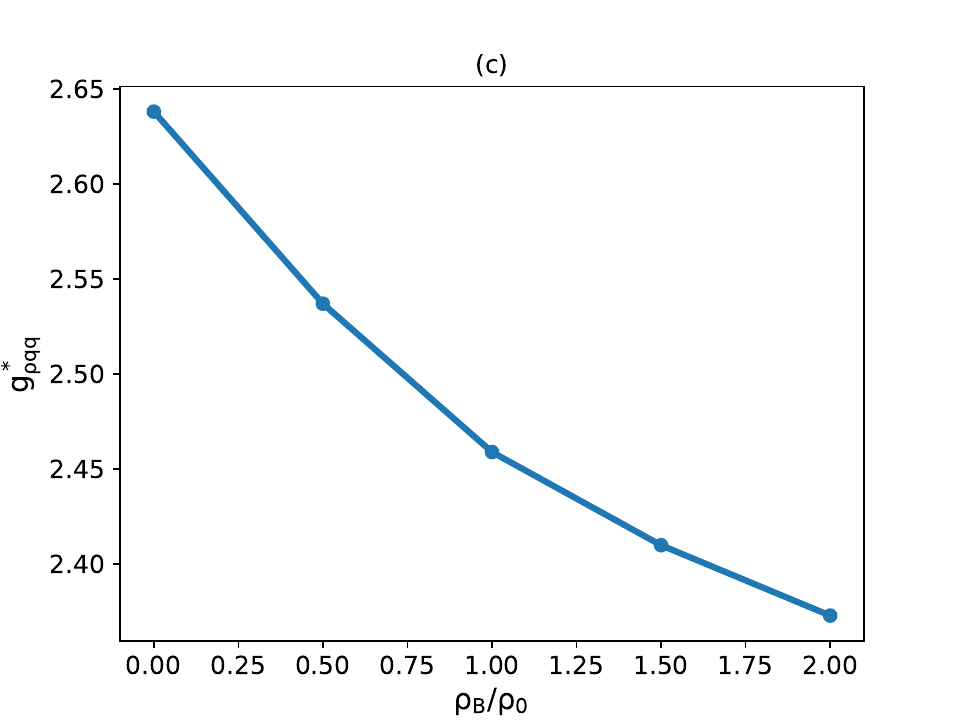} 
\includegraphics[width=0.49\columnwidth]{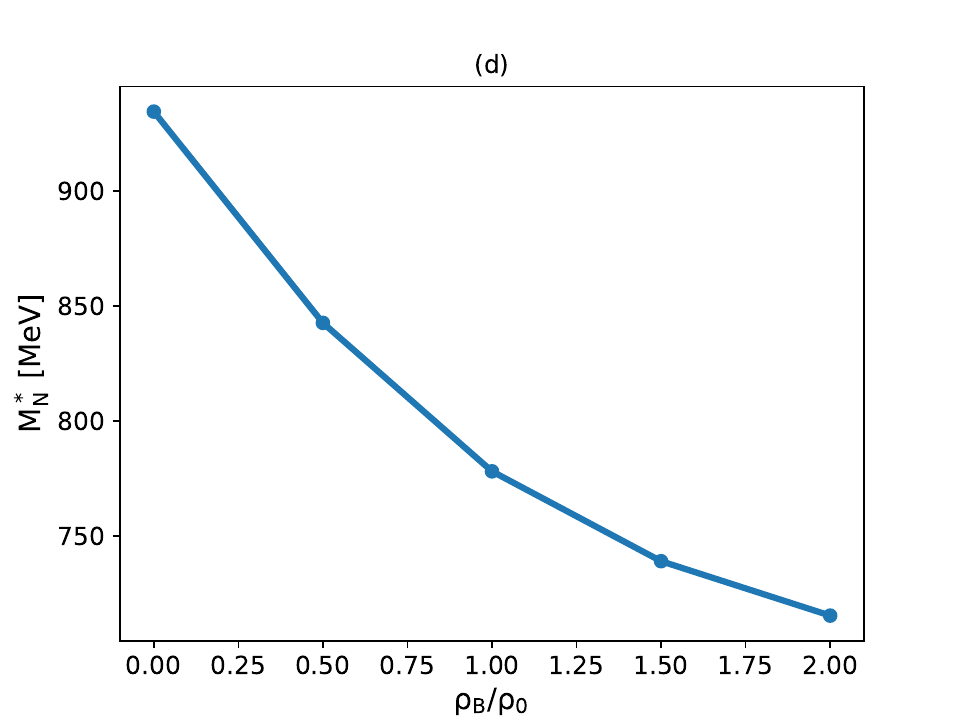} 
\caption{\label{fig1} Results for (a) $\rho$ meson effective mass, (b) dynamical quark effective
mass, (c) $\rho$-quark coupling constant,
and (d) nucleon effective mass as a function of $\rho_B/\rho_0$,
where $\rho_B$ and $\rho_0\,\, (=0.16\,\, \text{fm}^{-3})$
are respectively the baryon and saturation densities. Note that the in-medium $\rho$ meson and
dynamical effective quark masses represented by the filled asterisk (red) data are taken
from Ref.~\cite{deMelo:2018hfw}. Note that, as mentioned in the text, the results of Ref.~\cite{deMelo:2018hfw} used $\rho_0 = 0.15$ fm$^{-3}$ and thus, for an exact comparison with the present results ($\rho_0 = 0.16$ fm$^{-3}$), we adjust the horizontal values of the filled asterisk (red) data multiplying by a factor of (0.15/0.16) = 0.9375. Additionally, the ''hybrid'' model of the $\rho$-meson in Ref.~\cite{deMelo:2018hfw} can form
the bound state $\rho$ meson up to around $\rho_B/\rho_0 \simeq 0.9 \times 0.9375 = 0.84375$
in the present $\rho_B/\rho_0$ units. Thus, for larger nuclear matter densities, the in-medium $\rho$ meson properties could not be calculated within the ''hybrid model'' in
Ref.~\cite{deMelo:2018hfw}.}
\end{figure*}

Results for the $\rho$ meson properties in nuclear medium are shown in
Fig.~\ref{fig1}(a)-(d). Evaluating
the pole position of the $t$-matrix in Eq.~(\ref{eq:NJL5}), we obtain the $\rho$ meson mass in
nuclear medium, as depicted in Fig.~\ref{fig1}(a). It shows that the $\rho$ meson mass decreases as
the nuclear matter density increases. A similar behavior was found in
Refs.~\cite{deMelo:2018hfw,Shakin:1993ta}, represented by the filled asterisk (red) data,
where the in-medium $\rho$ meson mass also decreases as the nuclear matter density increases.
However, the magnitude is significantly different, in particular at $\rho_B /\rho_0 \gtrsim$ 0.5.
It is worth noting that in Ref.~\cite{deMelo:2018hfw}, they used $\rho_0$ is 0.15 fm$^{-3}$, which is different from the present work, $\rho_0 = 0.16$ fm$^{-3}$.
The difference in the $\rho$ meson mass in the nuclear medium is more pronounced at
higher densities due to the strong attractive scalar potential in the QMC model, as clearly
indicated in Fig.~\ref{fig1}(a).
We find that the $\rho$ mass reduction relative to that in free
space is approximately 10\% at normal nuclear matter density $\rho_B/\rho_0 =1.00$, which also may
correspond to the amount of the chiral symmetry restoration. Our result is consistent with
that obtained by the QCDSR of Ref.~\cite{Hatsuda:1991ez} and a very recent analysis in the IQCDSR of
Ref.~\cite{Mutuk:2025lak}, which are about (10--20)\% in the $\rho$ mass reduction. However, our result in $\rho$ mass reduction is rather different from that obtained in
Ref.~\cite{deMelo:2018hfw}, which is approximated by 33\%. The difference is probably related to a larger constituent light quark mass $m_q$ used in the model of Ref.~\cite{deMelo:2018hfw} (See Table I of the reference), where this  value of $m_q = 430$
MeV was set to be able to have the
''bound $\rho$ meson state", yielding the larger $g_\sigma$ and
$g_\omega$ coupling constant values to reproduce the empirically extracted nuclear matter
saturation properties in the QMC model.

In Fig.~\ref{fig1}(b), we also show the dynamical quark effective mass in symmetric nuclear matter calculated in
the NJL model that was described in Sec.~\ref{sec:NMNJL}. It is worth noting that this in-medium dynamical
quark mass was used as input to calculate the $\rho$ meson mass in symmetric nuclear matter
within the NJL model. As in the features of the in-medium $\rho$ meson mass, the dynamical quark
mass also decreases with increasing nuclear matter density.
The trend of our results for the $\rho$ meson and dynamical quark
masses is consistent with other theoretical predictions~\cite{deMelo:2018hfw},
but different in magnitude, as clearly indicated in Fig.~\ref{fig1}(b).
We also compare our present result with that
calculated in the QMC model reported in Ref.~\cite{deMelo:2018hfw}.
We find that the differences are
more significant at higher densities. Again, these differences are expected due to a strong scalar
attractive potential originated from the QMC model.
We find that the reduction of the dynamical quark effective mass at
$\rho_B/\rho_0 = 1.0$ relative to that in free space (vacuum) is approximately 18\% (72 MeV). The reduction of the constituent quark mass at $\rho_B/\rho_0 =0.9$ obtained in Ref.~\cite{deMelo:2018hfw} is 40\% (170 MeV), which is larger than our result. Note that to be exactly comparable with the present unit scale of $\rho_0 = $ 0.16 fm$^{-3}$, the nuclear matter density has to be adjusted by a factor of $0.9375$, for example, $\rho_B/\rho_0 =0.9 \times (0.15/0.16) = 0.9 \times 0.9375 = 0.84375$, as shown in Figs.~\ref{fig1}(a)-(b).

Next, we show the results of the $\rho$-quark coupling constant for different nuclear matter
densities in Fig.~\ref{fig1}(c).
It is worth noting that the $\rho$-quark coupling constant is related to the
$\rho$ meson wave function renormalization constants, which are given in
Eqs.~(\ref{eq:NJL6}) and~(\ref{eq:NJL6b}).
We find that the $\rho$-quark coupling constant decreases when the nuclear matter
density increases. Furthermore, we found that the reduction of the $\rho$-quark coupling
constant at normal nuclear matter density relative to that in free space is about 7\%.

Figure~\ref{fig1}(d) shows the results of the nucleon effective mass
as a function of the nuclear matter
density. The nucleon effective mass is calculated in the NJL nuclear matter model described in
Sec.~\ref{sec:NMNJL}, which shows the unique features of the NJL model approach computed at
the quark level. As a result, it shows that, as the nuclear matter density increases, the nucleon
effective mass decreases, which is consistent with other theoretical
predictions~\cite{deMelo:2018hfw,Zhu:2018vwn}. The reduction of the nucleon effective mass at normal
nuclear matter density relative to that in free space is about 17\% (156 MeV). We note that the NJL
model approach to nuclear matter reproduces the (negative of) energy for nuclear matter
at saturation density $\rho_0 =$ 0.16 fm$^{-3}$, as explained in
Ref.~\cite{Bentz:2001vc}. This indicates that the NJL model description of nuclear matter guarantees the nuclear matter stability, as empirically proven.

\subsection{In-medium $\rho$ meson EMFFs}
Before presenting the results for the charge, magnetic moment, and quadrupole form factors of the
$\rho^+$ meson in symmetric nuclear matter as well as in free space, we explore first the results
of the BSE dressed up-quark form
factors\footnote{In Ref.~\cite{Carrillo-Serrano:2015uca}, $F_{\mathrm{1U}}^* (Q^2)$ and
$F_{\mathrm{1D}}^* (Q^2)$ are usually written as $F_{\mathrm{1U}}^{* \mathrm{BSE}} (Q^2)$ and
$F_{\mathrm{1D}}^{* \mathrm{BSE}}(Q^2)$, respectively, where the BSE superscript indicates
that the form factors are uniquely
determined in the BSE. In this work, we suppress it for simplicity.}
$F_{\mathrm{1U}}^*(Q^2)$ and down-quark (antidown-quark can be obtained by multiplying by a factor of $-1$) form
factors, $F_{\mathrm{1D}}^* (Q^2)$, where the expression of this inhomogeneous
BSE dressed quark form factors are formulated in Eq.~(\ref{eq:vff4}).
\begin{figure}[t!]
\centering
\includegraphics[width=0.49\columnwidth]{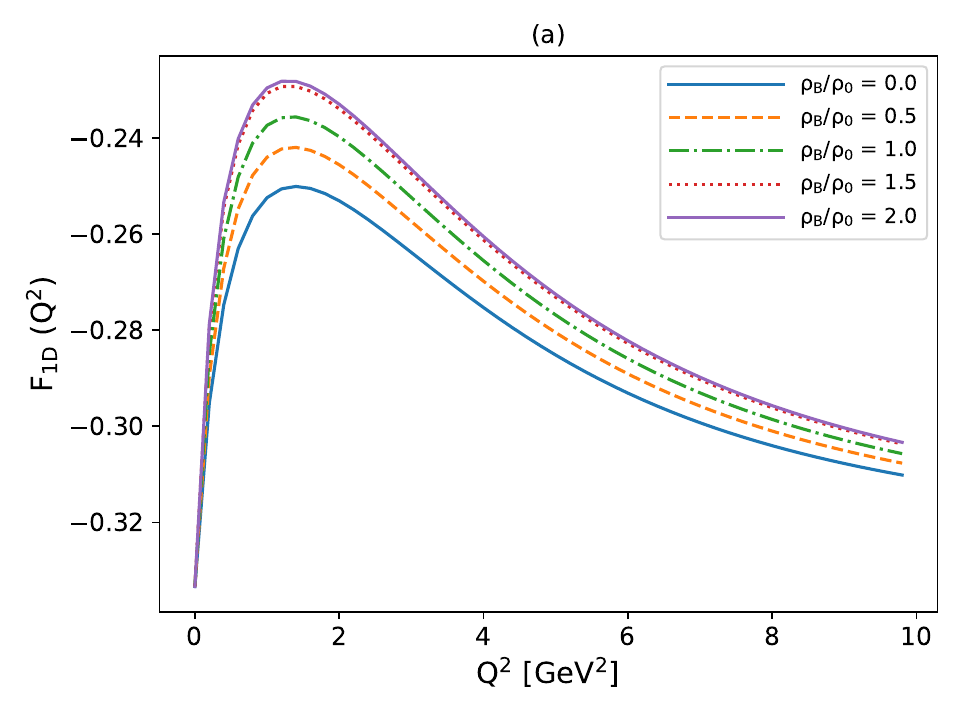} 
\includegraphics[width=0.49\columnwidth]{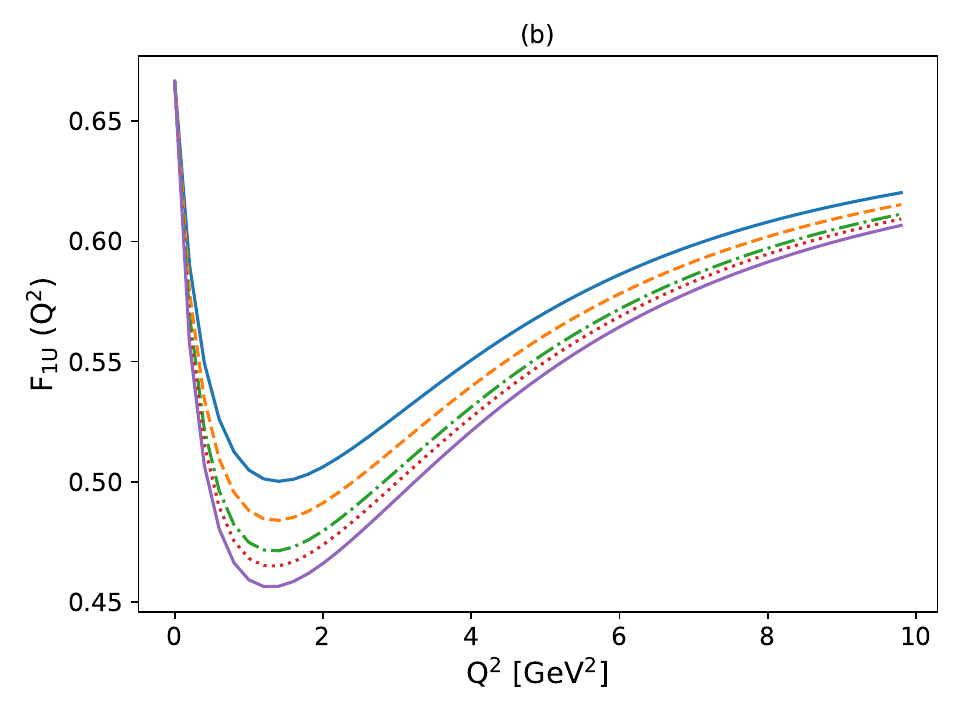} 
\caption{\label{fig2} Results for the BSE dressed quark form factors for different nuclear matter densities,
(a) BSE dressed form factor of the down quark, $F_{\mathrm{1D}}^*(Q^2)$,
and (b) BSE dressed form factor of the up quark,
$F_{\mathrm{1U}}^* (Q^2)$ for different densities as a function of $Q^2$.}
\end{figure}

Results for BSE dressed down-quark form factor $F_{\mathrm{1D}}^* (Q^2)$ for different nuclear
matter densities
as a function of $Q^2$ are shown in Fig.~\ref{fig2}(a), where we can see that
$F_{\mathrm{1D}}^*(Q^2)$ increases as the nuclear matter density increases. At $Q^2 =0$ we have
$F_{\mathrm{1D}}^*(Q^2=0) = e_d = -1/3$, where $e_d$ stands for the
electric charge of the down quark. It is worth noting that this normalization value at $Q^2 =0$ is
normalized not only in free space but also in nuclear matter.
Also, we find that $F_{\mathrm{1D}}^*(Q^2)$ increases up to
around $Q^2 \simeq $ 2 GeV$^2$ and begins to decrease as the $Q^2$ increases. It is worth noting
that, in this work, we do not consider the pion loops contributions in the $\rho$ meson EMFFs
as in Refs.~\cite{Carrillo-Serrano:2015uca,Cloet:2014rja}, since we focus on the nuclear
medium effects on the $\rho$ meson EMFFs. The pion loop contribution to the $\rho$ meson EMFFs in nuclear matter is expected to be insignificant, because the physical
pion mass remains nearly unchanged in the nuclear medium, as indicated in
Ref.~\cite{Gifari:2024ssz} (and references therein).

As $Q^2 \rightarrow \infty$ in free space, the $F_{\mathrm{1D}}^* (Q^2)$ does
not go to zero, but it behaves as $F_{\mathrm{1D}}^* (Q^2) \simeq e_d$.
A similar feature is captured in the
$F_{\mathrm{1D}}^* (Q^2)$ in nuclear matter, where it also gives $F_{\mathrm{1D}}^* (Q^2) \simeq
e_d$ at $Q^2 \rightarrow \infty$. Such a behavior at larger $Q^2$ is consistent with the expectation
results from asymptotic freedom of QCD~\cite{Cloet:2014rja}.
We also compute the BSE dressed form factors for the up quark. The results for the
$F_{\mathrm{1U}}^* (Q^2)$ as
a function of $Q^2$ for different nuclear matter densities are illustrated in Fig.~\ref{fig2}(b). It
is clearly shown that the behavior of $F_{\mathrm{1U}}^* (Q^2)$ is rather different from that
for $F_{\mathrm{1D}}^* (Q^2)$, where the $F_{\mathrm{1U}}^* (Q^2)$ decreases as the nuclear matter
density increases. As a function of $Q^2$, the $F_{\mathrm{1U}}^* (Q^2)$ form factor decreases up to
$Q^2 \simeq 2$ GeV$^2$, and then increases at higher $Q^2$, exhibiting an opposite trend to that of
$F_{\mathrm{1D}}^*(Q^2)$.
It is worth noting that if the absolute value is taken, or the sign is changed for the dressed $D$
quark form factor as $-F_{\mathrm{1D}}^* (Q^2)$, the global
behavior looks similar except for the magnitude difference with $F_{\mathrm{1U}}^* (Q^2)$.

Figure~\ref{fig2}(b) also shows that  $F_{\mathrm{1U}}^* (Q^2=0) = e_u = 2/3$, which is the value of the up quark electric charge. This value is reproduced in free space (vacuum) and nuclear matter, as
clearly shown in Fig.~\ref{fig2}(b) at $Q^2 =0$. For larger $Q^2$ (goes to infinity), we find that
$F_{\mathrm{1U}}^* (Q^2\rightarrow \infty) \simeq e_u$, which is consistent with what we expect as the QCD behavior in the asymptotic regime.
\begin{figure}[ht]
	\centering
    \includegraphics[width=0.49\columnwidth]{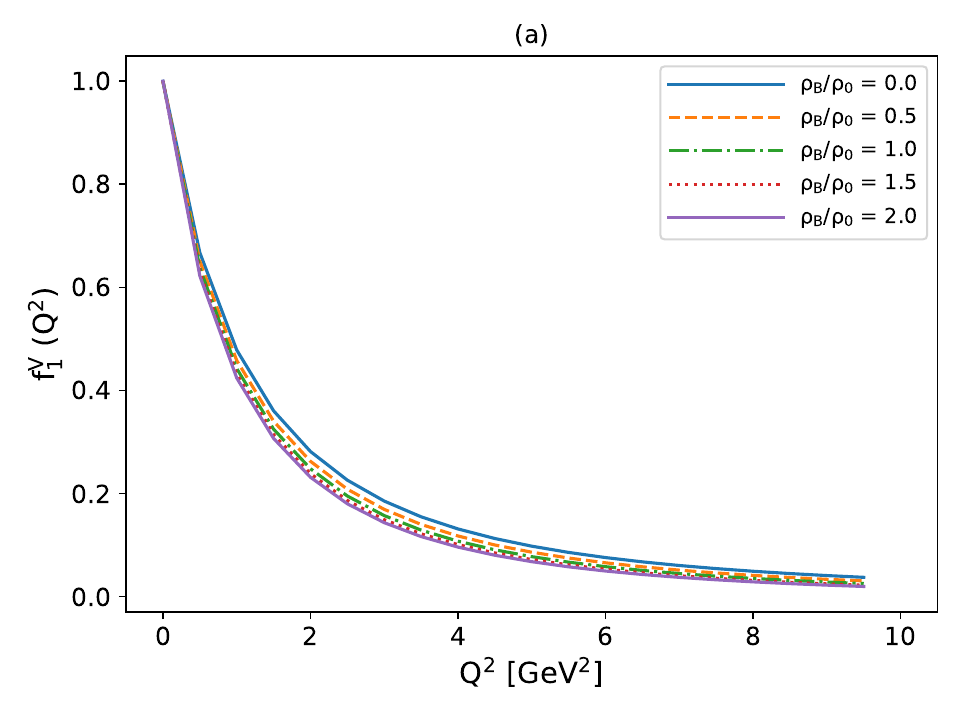} 
    \includegraphics[width=0.49\columnwidth]{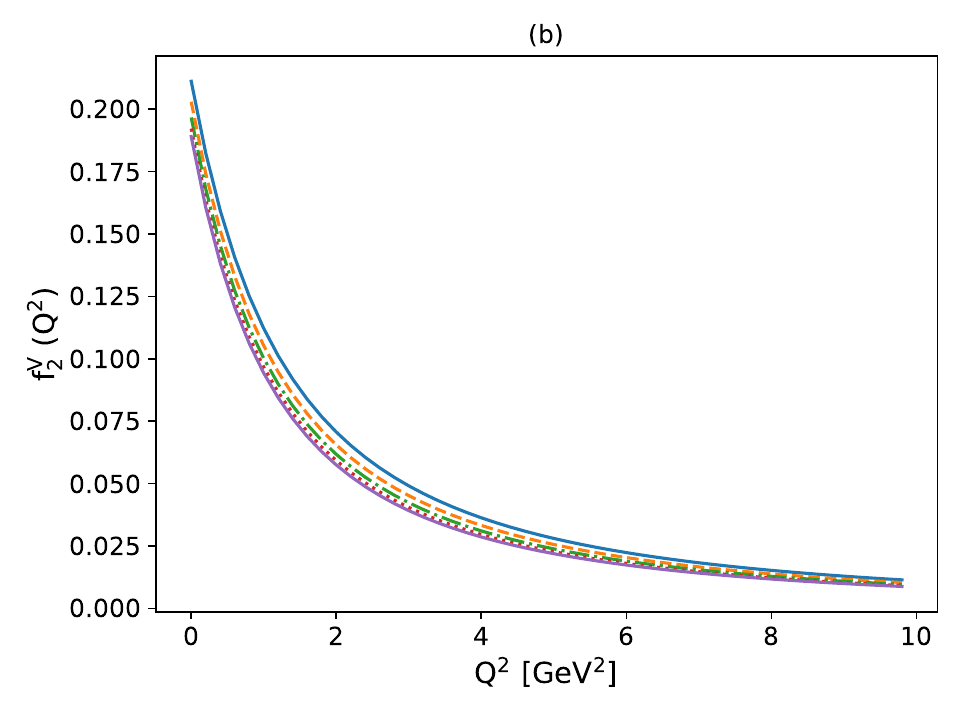} 
    \includegraphics[width=0.49\columnwidth]{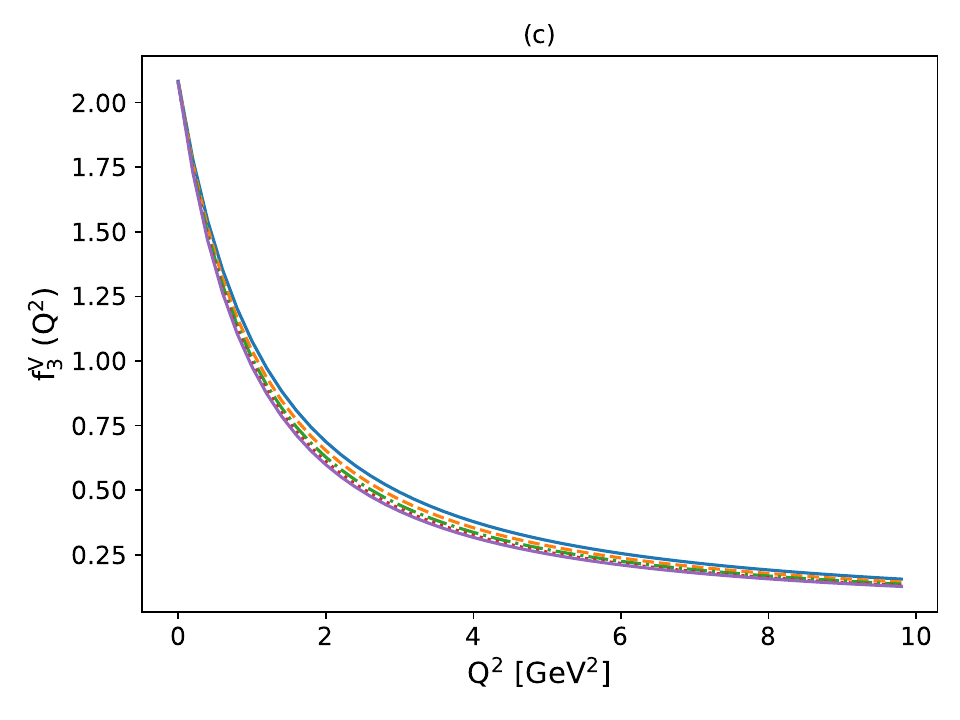} 
  \caption{\label{fig3} Results of the vector body form factors, (a) $f_{1}^{* V} (Q^2)$, (b)
$f_{2}^{* V}
  (Q^2)$, and (c) $f_{3}^{* V} (Q^2)$ for different nuclear matter densities as a function of $Q^2$.}
\end{figure}

In computing the $\rho$ meson EMFFs in a nuclear medium,
the medium effects are also expected to
influence the vector body form factors, as formulated in Eqs.~(\ref{eq:vff3})-(\ref{eq:vff3b}).
Results for the vector body form factors for different nuclear matter densities as a function of
$Q^2$ are depicted in Fig.~\ref{fig3}(a)-(c). It is worth noting that, in addition to
the vector body form factors, in general, there also exist the tensor body form factors,
which come from the tensor part of
${\sigma^{\mu \nu} q_\nu}/{2M}$. However, in the present work, the contributions from the
tensor body form factors vanish, because $F_{\mathrm{2Q}}^*(Q^2)$ is zero, with the dressed quark
$Q =(U,D)$.

Figure~\ref{fig3}(a) shows that the $f_{1}^{* V} (Q^2)$ decreases as the nuclear matter density
$\rho_B/\rho_0$ and $Q^2$ increase.
Note that the $f_{1}^{V} (Q^2)$ is unity at $Q^2=0$ in free space as well as in nuclear
matter. This corresponds that $f_{1}^{* V} (Q^2)$ satisfies the charge conservation. Such a
normalization condition is also found in Refs.~\cite{Carrillo-Serrano:2015uca,Cloet:2014rja}
for a free space case.
Furthermore, Fig.~\ref{fig3}(a) also exhibits that the change of $f_1^{* V} (Q^2)$ is more
pronounced at higher densities.

In addition to the results for $f_{1}^{* V} (Q^2)$, we also show the results for the vector body
form factors
$f_2^{* V} (Q^2)$ for different nuclear matter densities as a function of $Q^2$, as illustrated in
Fig.~\ref{fig3}(b). We find that the $f_2^{* V} (Q^2)$ also decreases as the $\rho_B/\rho_0$ and
$Q^2$ increase. Also, the values of $f_2^{* V} (Q^2=0)$ decrease as the nuclear matter density
increases. The reduction of $f_2^{* V} (Q^2 =0)$ at $\rho_B/\rho_0 = 1.0$ is 6.87\% in comparison to
that for the free space. At higher density $\rho_B /\rho_0 = 2.0$, the decrease of $f_2^{* V} (Q^2
=0)$ is 10.49\%.
In Fig.~\ref{fig3}(c), we show our results for $f_3^{* V} (Q^2)$ as a function of $Q^2$ for
different
densities. In free space, $f_3^{* V} (Q^2=0) =$ 2.09, which is the magnetic moment of the $\rho^+$ meson in the units of $e/(2m_\rho)$, where $e=1$ and $m_\rho$
are the natural units of
the positron charge $e$ and
$\rho$ meson mass, respectively. This value is rather larger than the canonical value~\cite{Cloet:2014rja}.
In symmetric nuclear matter, the magnetic moment
decreases as the nuclear matter density increases. For example, the values of
$f_3^{* V} (Q^2 = 0) = 2.079$ at $\rho_B/\rho_0 = 1.0$.
For $\rho_B/\rho_0 = 2.0$, we have $f_3^{*V} (Q^2 =0) = 2.077$.
\begin{figure}[ht]
	\centering
    \includegraphics[width=0.49\columnwidth]{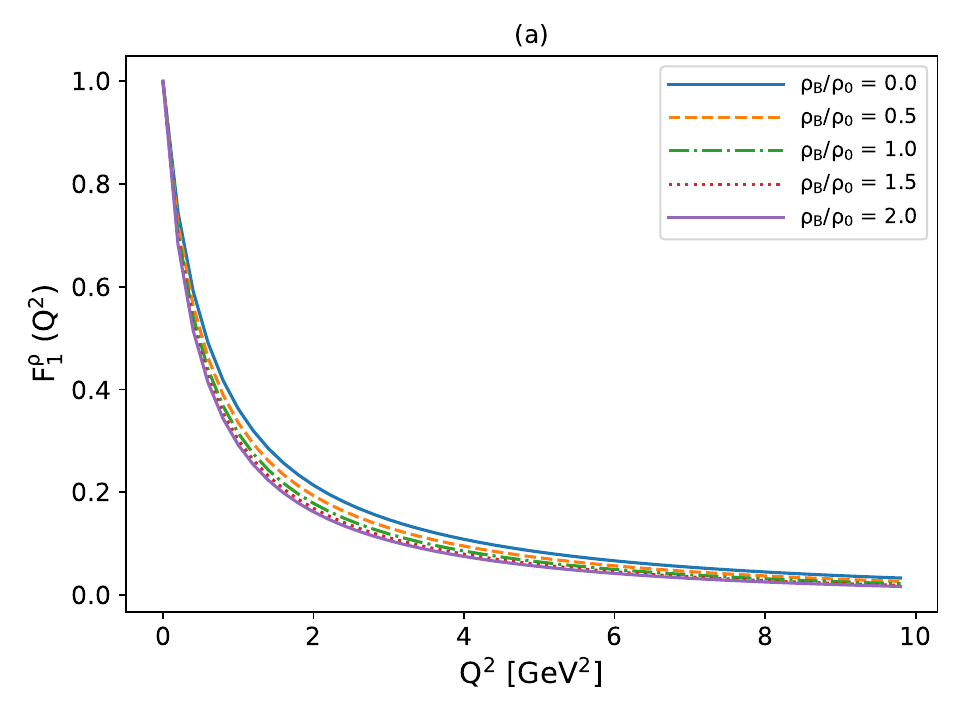} 
    \includegraphics[width=0.49\columnwidth]{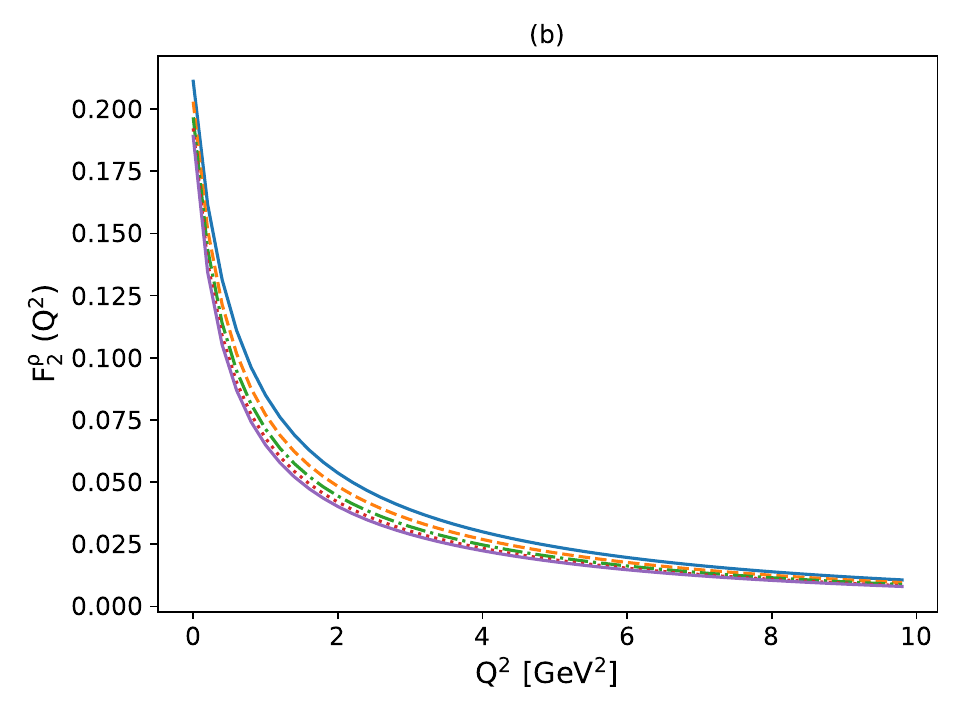} 
    \includegraphics[width=0.49\columnwidth]{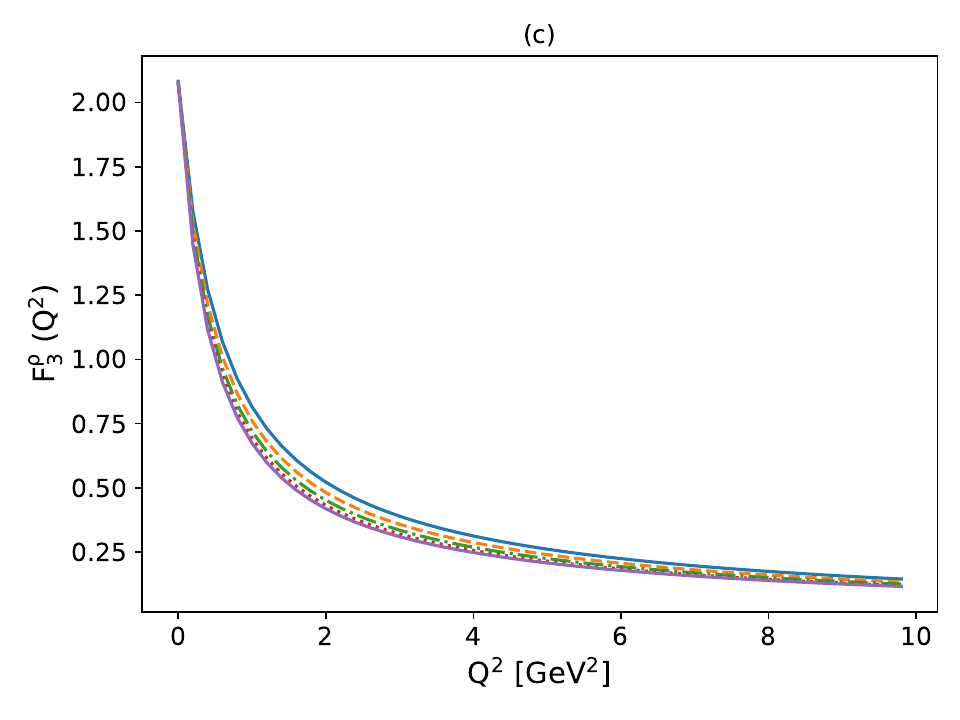} 
  \caption{\label{fig4} Results of the $\rho^+$ meson EMFFs for different densities
(a) $F_{1}^{* \rho} (Q^2)$, (b) $F_2^{* \rho} (Q^2)$, and (c) $F_{3}^{* \rho} (Q^2)$ for different nuclear matter
densities as a function of $Q^2$.}
\end{figure}
Now we turn to discuss the results of the $\rho^+$ meson EMFFs, $F_{1}^{\rho} (Q^2)$,
$F_{2}^{\rho} (Q^2)$, and $F_{3}^{\rho} (Q^2)$ in free space as well as in symmetric
nuclear matter as depicted in Figs.~\ref{fig4}(a)-(c).
Results of $F_{1}^{* \rho} (Q^2)$ for different nuclear matter densities as a function of $Q^2$ are
illustrated in Fig.~\ref{fig4}(a). At $\rho_B/\rho_0 = 0.0$ (free space), our result on $F_1^\rho
(Q^2)$ is consistent with that obtained in Ref.~\cite{Cloet:2014rja}. As the nuclear matter density
and $Q^2$ increase, $F_1^{* \rho} (Q^2)$ decreases. It is worth noting that $F_{1}^{* \rho} (Q^2)$
always satisfies the normalization condition in free space and nuclear matter,
i.e., $F_{1}^{\rho} (Q^2 =0) = 1.0$.

In Fig.~\ref{fig4}(b), we show the results of the $F_2^{* \rho} (Q^2)$ for different densities as a
function of
$Q^2$. We find that the results for $F_2^{* \rho} (Q^2)$ decrease as the nuclear matter density
and $Q^2$ increase. The decrease of $F_2^{* \rho} (Q^2)$ is more pronounced at higher nuclear matter
densities. We also find that in free space, our results are consistent with those reported in
Ref.~\cite{Cloet:2014rja}.

Results for the $F_3^{* \rho} (Q^2)$ for different nuclear matter densities as a function of $Q^2$
are given in
Fig.~\ref{fig4}(c). Similar to the results of $F_{1}^{* \rho} (Q^2)$ and $F_2^{* \rho} (Q^2)$, we
find that the $F_3^{* \rho} (Q^2)$ decreases as the nuclear matter density $\rho_B/\rho_0$ and $Q^2$ increase. It is more significant at higher $\rho_B/\rho_0$. Again, our results of
$F_3^{ \rho}(Q^2)$ in free space are consistent with those obtained in Ref.~\cite{Cloet:2014rja}.
It is worth noting that in free space, $F_3^\rho (Q^2 =0)$ equals the magnetic moment of the $\rho$
meson. The decrease of $F_3^{* \rho} (Q^2 =0)$ at $\rho_B/\rho_0 = 1.0$ is approximately 0.15\% in
comparison to that for free space ($\rho_B/\rho_0 =0.0$), which is insignificant reduction. At
higher nuclear matter density, $\rho_B/\rho_0 = 2.0$,
we find a reduction of the $F_3^{* \rho} (Q^2=0)$ by
approximately 0.24\% in comparison to that in free space, and again, insignificant.
\begin{figure}[t]
\centering
\includegraphics[width=0.49\columnwidth]{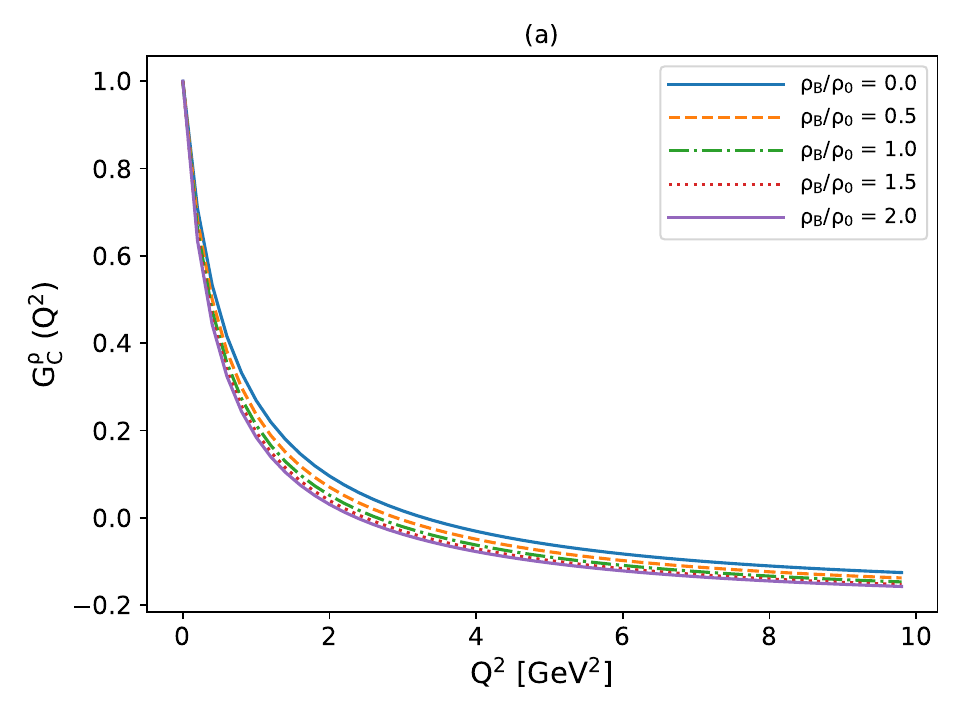} 
\includegraphics[width=0.49\columnwidth]{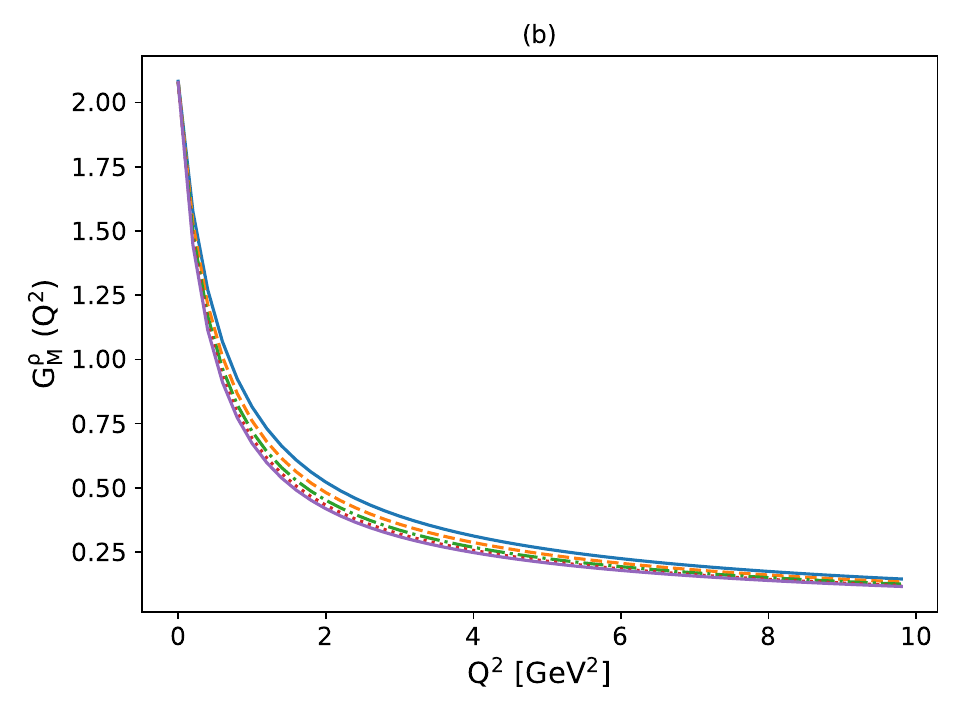} 
\includegraphics[width=0.49\columnwidth]{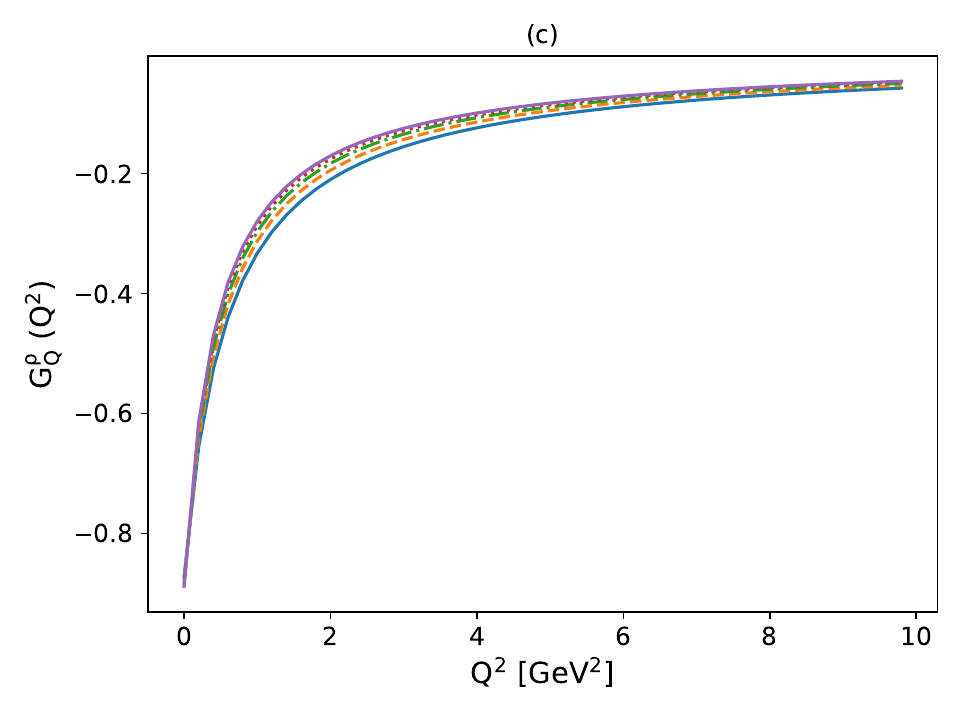} 
\includegraphics[width=0.49\columnwidth]{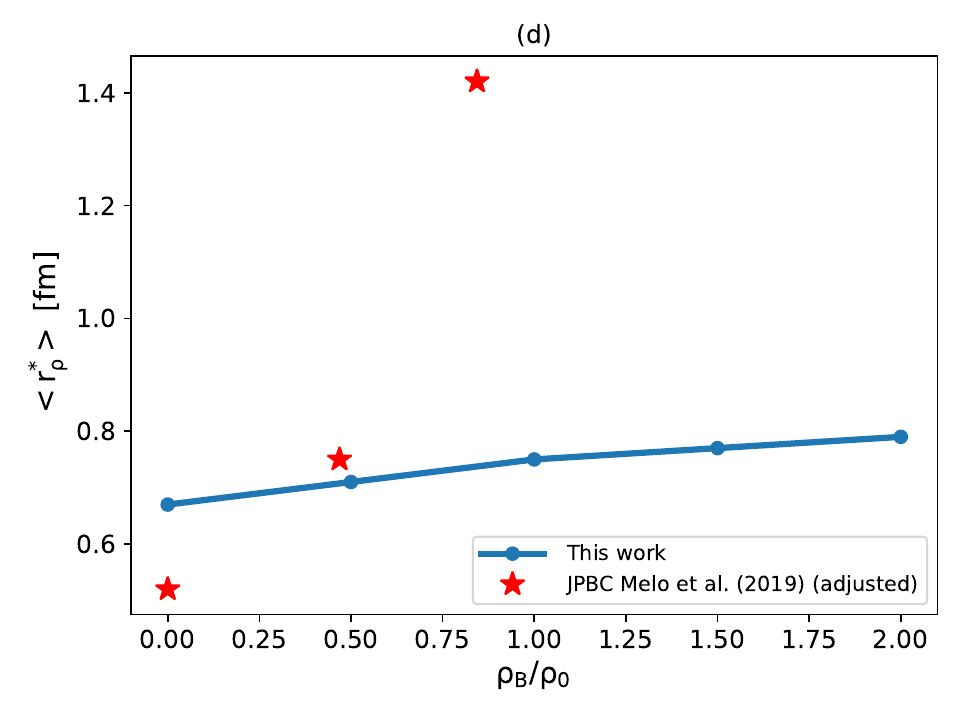}
\caption{\label{fig5} Results for the $\rho^+$ meson EM properties, (a) the charge form factors, (b)
magnetic form factors, (c) quadrupole form factors, and (d) charge radius, $\big<r_\rho^* \big> \equiv \big< r_C^{*2} \big>^{1/2}$, as a function of
$\rho_B/\rho_0$. See also the caption of Fig.~\ref{fig1} for the note on the asterisk (red) data.}

\end{figure}

We are now in a position to present the in-medium $\rho^+$ meson EMFFs: the charge form factor
$G_C^{\rho} (Q^2)$ in units of the positron charge $e$, the magnetic form factor
$G_M^{\rho} (Q^2)$ in units of $e/(2m_\rho)$, and the quadrupole form factor $G_Q^{ \rho} (Q^2)$
in units of $e/m_\rho^2$, both in free space and in nuclear matter.
Results for the $\rho^+$ meson charge form factor for different
nuclear matter densities as a function of $Q^2$ are illustrated in Fig.~\ref{fig5}(a).
We find that the values of $G_C^\rho (Q^2)$ in free space are positive
up to $Q^2 \leq 3.2$ GeV$^2$. At
$Q^2 \geq 3.4$ GeV$^2$, the values of $G_C^\rho (Q^2)$ become negative.
At $Q^2 =0$, $G_C^\rho (Q^2)$ satisfies the charge normalization (charge conservation), namely,
$G_C^\rho (0) = 1.0$. This occurs not only in free space but also in a nuclear
medium. At $\rho_B/\rho_0 = 1.0$, the
positive values of $G_C^{* \rho} (Q^2)$ slightly move up to $Q^2 \leq 2.6$ GeV$^2$, and for
$Q^2 \geq 2.8$ GeV$^2$, the values of $G_C^{* \rho} (Q^2)$
become negative, which starts smaller $Q^2$ value than that for
free space. At higher nuclear matter density $\rho_B/\rho_0 = 2.0$,
the values of $G_C^{* \rho} (Q^2)$ become positive for $Q^2 \leq 2.2$ GeV$^2$,
while $G_C^{* \rho} (Q^2)$ becomes negative at $Q^2 \geq 2.4$ GeV$^2$.
Our result of $G_C^{\rho} (Q^2)$ in free space is consistent with
those reported in Ref.~\cite{Cloet:2014rja}.

Let us now discuss our results for the $G_C^{* \rho} (Q^2)$ in the nuclear medium.
Our results are consistent with those obtained in Ref.~\cite{deMelo:2018hfw}.
However, the differences are in the $Q^2$ values for
crossing negative-positive values and the magnitudes.
The results reported in Ref.~\cite{deMelo:2018hfw} indicated that $G_C^{* \rho} (Q^2)$ in nuclear matter
is softer compared to our results. This is indicated by the onset of
the negative values in $G_C^{*\rho} (Q^2)$ occurring at a lower $Q^2$, i.e., at a smaller $Q^2$ value.

Next, in Fig.~\ref{fig5}(b), we show our predictions for $\rho^+$ meson magnetic moment
form factors, $G_M^{* \rho} (Q^2)$ for different values of $\rho_B/\rho_0$ and $Q^2$.
We find that the $G_M^{* \rho} (Q^2)$ decreases as the $\rho_B/\rho_0$ and $Q^2$ increase.
This behavior of $G_M^{* \rho} (Q^2)$ is consistent with the trend reported in
Ref.~\cite{deMelo:2018hfw}, although the magnitudes differ.
Note that the $G_M^{* \rho} (Q^2)$ at $Q^2 = 0$ equals to the $\rho^+$ meson magnetic moment,
$\mu_\rho^*$. In this work, we follow the formula of Ref.~\cite{Carrillo-Serrano:2015uca} in
free space, where $\mu_\rho = G_M (0) \frac{M_N}{m_\rho}$. In the nuclear medium, the expression of
$\mu_\rho$ is defined as $\mu^*_\rho = G_M^* (0) \frac{M_N^*}{m_\rho^*}$.
The values of $m_\rho^*$ and $M_N^*$ for various nuclear matter densities are given in Table~\ref{tab:NJL2}.
We find that the value of $G_M^{\rho} (Q^2)$ in free
space at $Q^2 = 0$ is 2.082, in agreement with the value obtained in Ref.~\cite{Cloet:2014rja},
while at $\rho_B/\rho_0 =1.0$, we obtain $G_M^{* \rho} (0) =$ 2.079,
which is very similar magnitude to that in free space. At higher density $\rho_B/\rho_0 = 2.0$, the value of $G_M^{* \rho} (0)$ is
2.077. The value of $G_M^{* \rho} (0)$ is 2.081 at $\rho_B/\rho_0 =0.5$,
which is smaller than the result obtained in Ref.~\cite{deMelo:2018hfw},
i.e., $G_M^{* \rho} (0) = 2.18$ at $\rho_B/\rho_0=0.5$.

Results for the quadrupole moment form factors of $\rho^+$ meson $G_Q^{* \rho} (Q^2)$
for different nuclear matter densities as a function of $Q^2$ are illustrated in
Fig.~\ref{fig5}(c). We find that the
$G_Q^{* \rho} (Q^2)$ increases as the nuclear matter density and $Q^2$ increase.
In free space, our result is consistent with that obtained in
Ref.~\cite{Carrillo-Serrano:2015uca,Cloet:2014rja}.
Also, we obtain the same values of $G_Q^\rho (Q^2)$ at $Q^2 =0$ with those reported in
Refs.~\cite{Carrillo-Serrano:2015uca,Cloet:2014rja}, i.e., $G_Q^\rho (0) = -0.87$ in free space.
However, this value is rather different from that obtained in Ref.~\cite{deMelo:2018hfw} in free
space and also the value in nuclear matter.
The difference is not only found in the values of $G_Q^{* \rho} (0)$, but
also in the $Q^2$ dependence of $G_Q^\rho (Q^2)$ for specific nuclear matter densities as
shown in the lower-left panel of Fig.~5 in Ref.~\cite{deMelo:2018hfw},
where there is zero crossings in $G_Q^{* \rho} (Q^2)$ for $\rho_B/\rho_0 =0.00,
0.25,$ and $0.50$ at low $Q^2$. Such behavior is not found in
our results for the $G_Q^{* \rho} (Q^2)$, as clearly shown in Fig.~\ref{fig5}(c).
We find that the values of the $G_Q^{* \rho} (0)$ for $\rho_B/\rho_0 = 0.5, 1.0, 1.5,$ and $2.0$
are $-0.878, -0.883, -0.886,$ and $-0.888$, respectively.

In Fig.~\ref{fig5} (d), we show the results for the $\rho^+$ meson in-medium charge radius
$\big<r_\rho^*\big> \equiv \big< r_C^{*2} \big>^{1/2}$ as a function of $\rho_B/\rho_0$.
We find that the charge radius increases as the nuclear
matter density increases. In free space, it is consistent with the value obtained in
Ref.~\cite{Cloet:2014rja}, i.e., $\big< r_\rho \big> = 0.67 $ fm.
However, this is rather different from that obtained in Ref.~\cite{deMelo:2018hfw},
i.e., $\big< r_\rho \big> = 0.52$ fm, as also clearly indicated in Fig.~\ref{fig5}(d).
Furthermore, our result in free space differs significantly
from that of Ref.~\cite{Carrillo-Serrano:2015uca}, where it was reported $\langle r_\rho \rangle = 0.82$ fm.
This discrepancy arises from the absence of pion loop contributions in the present work.
Figure~\ref{fig5}(d) also shows a comparison of our charge radius with that of
Ref.~\cite{deMelo:2018hfw}. The charge radius of Ref.~\cite{deMelo:2018hfw} at
$\rho_B/\rho_0 = 0.50$ (red filled asterisks) are close to our results (blue solid line). However, the results for
$\big< r_\rho^* \big> $ are quite different from our results, as one can clearly see
at $\rho_B/\rho_0 = (0.0,$ and $0.90$). Note that, as already explained earlier, here we emphasize again that the
$\rho_B/\rho_0 = 0.9$ corresponds to 0.84375 in the present study, as indicated in Fig.~\ref{fig5}(d).
For convenience, the values of the charge radius, quadrupole moment, and
magnetic moment of the $\rho$ meson at various densities are listed in Table~\ref{tab:NJL2}.
\begin{figure}[t]
\centering
\includegraphics[width=0.49\columnwidth]{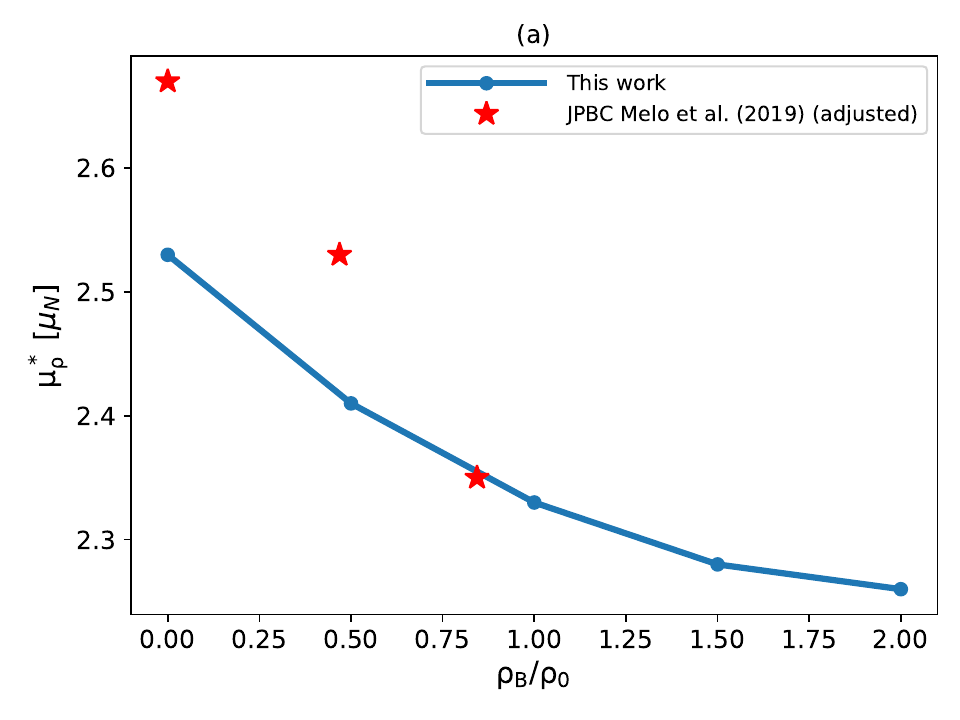} 
\includegraphics[width=0.49\columnwidth]{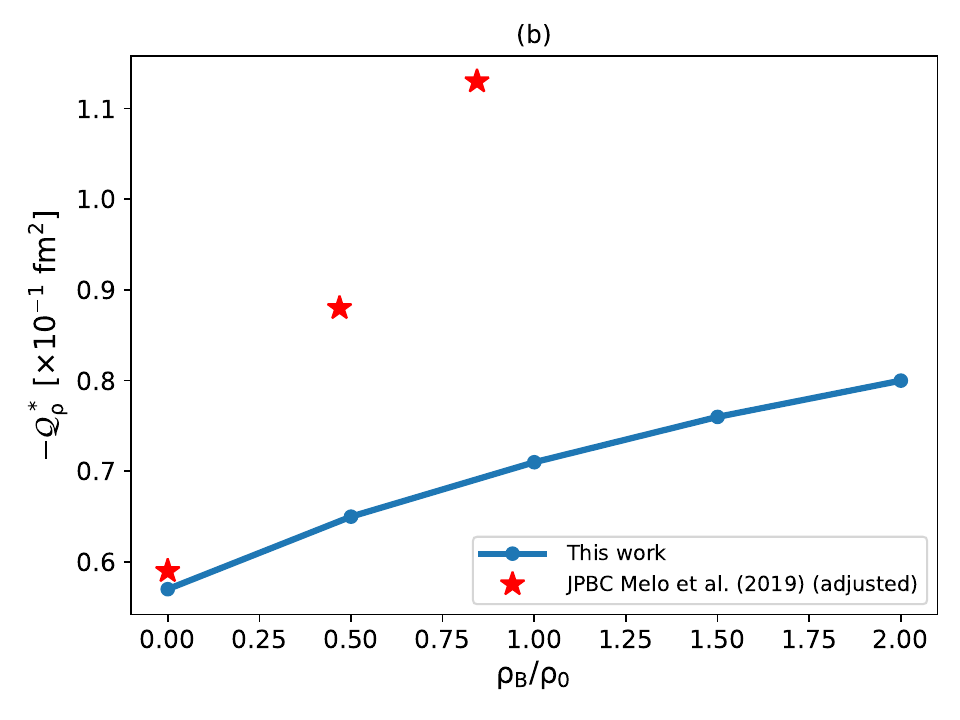} 
\caption{\label{fig6} Results for (a) magnetic moment, and (b) quadrupole moment of $\rho^+$ meson as a function of $\rho_B/\rho_0$. See also the caption of Fig.~\ref{fig1} for the note on the filled asterisk (red) data.}
\end{figure}

Finally, we show the results of the $\rho^+$ meson magnetic and quadrupole moments for
different nuclear matter densities, respectively, compared with those reported in
Ref.~\cite{deMelo:2018hfw}, as depicted in Fig.~\ref{fig6}(a)-(b).
In Fig.~\ref{fig6}(a), we show the result for the magnetic moment.
We find that the $\rho^+$ meson magnetic moment decreases as the nuclear matter density increases.
In addition, we find that our results for the magnetic moment are rather close to those
obtained in Ref.~\cite{deMelo:2018hfw}.
The behavior of $\mu_\rho^*$ exhibits a similar trend to our
prediction,
as in Fig.~\ref{fig6}(a), where the $\rho^+$ meson magnetic moment decreases with increasing
nuclear matter density. We find $\mu_\rho^* = 2.41 \mu_N$ at $\rho_B/\rho_0 = 0.5$, whereas
Ref.~\cite{deMelo:2018hfw} reports $\mu_\rho^* = 2.53 \mu_N$ at the same density.

In Fig.~\ref{fig6}(b), we show the results for the quadrupole moment of
the $\rho^+$ meson as a function of the
nuclear matter densities. We find that the $\mathcal{Q}_\rho^*$ increases as the nuclear matter
density increases. Again, a similar trend is reported in Ref.~\cite{deMelo:2018hfw},
indicating the $\mathcal{Q}_\rho^*$ increases as the $\rho_B/\rho_0$ increases.
At $\rho_B/\rho_0 = 0.0$ (free space) they obtained $\mathcal{Q}_\rho^* = -0.059$ fm$^2$,
which is close to our free space value, $\mathcal{Q}_\rho^* = -0.057$ fm$^2$.
However, at $\rho_B/\rho_0 =0.5$ in their units with $\rho_0 = 0.15$ fm$^{-3}$, namely 0.46785 in the present units,
the values of $\mathcal{Q}_\rho^*$ is rather different. At this point, our calculation yields
$\mathcal{Q}_\rho^*=-0.065$ fm$^2$, whereas the value they obtained
is $\mathcal{Q}_\rho^* = -0.088$ fm$^2$, which is smaller than ours.
As  $\rho_B/\rho_0$ increases, the difference becomes more significant.

\section{Summary and Conclusion} \label{sec:summary}
To summarize, we have investigated the in-medium modifications of the electromagnetic properties
(form factors and charge radius) of the
$\rho$ meson within the framework of the Nambu--Jona-Lasino (NJL) model,
utilizing the Schwinger proper-time regularization scheme to simulate quark confinement.
The effect of the nuclear medium is consistently studied in the NJL model
at the quark level, where the chiral symmetry breaking and the partial restoration can be described
in the NJL model
\textit{via} the quark condensate appearing in the model.
Thus, we systematically investigate the properties and structure of
$\rho$ meson in free space and in symmetric nuclear matter, such as the effective mass,
in-medium dynamical quark mass, in-medium $\rho$-quark coupling constant,
and $\rho$ meson electromagnetic properties, namely charge form factor, charge radius,
magnetic moment, and quadrupole moment.

As a result, we have found that the $\rho$ meson effective mass decreases as the nuclear matter
density increases. This is consistent with that obtained in Ref.~\cite{deMelo:2018hfw}, although the magnitude is different.
We find the $\rho$ meson mass reduction at
$\rho_B/\rho_0 =1.0$ relative to that in free space is about 10\%,
which is consistent with that obtained in the QCD sun rule
(QCDSR) of Ref.~\cite{Hatsuda:1991ez},
and a very recent analysis in the inverse QCD sum rule (IQCDSR) of
Ref.~\cite{Mutuk:2025lak}, giving about (10--20)\%.
However, the reduction of the $m_\rho^*$ of Ref.~\cite{deMelo:2018hfw} is rather
different from that obtained in the present work, as clearly shown in
Fig.~\ref{fig1}(a), where in the former, the in-medium $\rho$ mass was calculated in the quark-meson coupling (QMC) model using a rather larger free space light quark mass value compared to that used in the standard QMC model.

A similar trend is found in the in-medium dynamical quark mass,
where the in-medium dynamical quark mass decreases as the nuclear matter density increases, as
clearly illustrated in Fig.~\ref{fig1}(b).
The in-medium constituent quark mass results reported in Ref.~\cite{deMelo:2018hfw}
exhibit a similar trend to our predictions, showing a decrease with increasing nuclear matter
density. However, their results show a more rapid decrease in the in-medium
constituent quark mass calculated by the QMC model
with a large free space light quark mass value.

We predict the in-medium $\rho$-meson-quark coupling constant in the nuclear medium. As
expected, the $\rho$-meson-quark coupling constant decreases as the nuclear matter density
increases. We also find that the reduction of the $\rho$-meson-quark coupling constant at
$\rho_B/\rho_0 = 1.0$ relative to that in free space is approximately 7\%.

We have presented that the Bethe-Salpeter equation (BSE) calculated
dressed quark electromagnetic form factors for different densities and $Q^2$.
We have found that $F_{\mathrm{1D}}^* (Q^2)$ increases up to $Q^2 \simeq 2$ GeV$^2$
and then continues to decrease up to higher $Q^2$.
However, $F_{\mathrm{1D}}^* (Q^2)$ does not go to zero.
It behaves as $F_{\mathrm{1D}}^* (Q^2) = e_d$ at $Q^2 \rightarrow \infty$,
which is consistent with the
expectation of QCD in the asymptotic regime.

We have also shown the results for the $F_{\mathrm{1U}}^* (Q^2)$,
which exhibits the decrease of $F_{\mathrm{1U}}^* (Q^2)$ up to $Q^2 \simeq 2$ GeV$^2$ and then
continues to increase as the $Q^2$ increases.
However, it decreases as the nuclear matter density increases.
It is worth noting that the opposite of $Q^2$
dependent behavior to that of $F_{\mathrm{1D}}^* (Q^2)$, is due to the opposite sign of the charge.
At $Q^2 \rightarrow \infty$, $F_{\mathrm{1U}}^* (Q^2) \simeq e_u$,
which is consistent with the QCD expectation in the asymptotic regime.

For the results of the vector body form factors, we have found that,
$f_1^{* V} (Q^2)$, $f_2^{* V} (Q^2)$, and $f_3^{* V} (Q^2)$ decrease
as the nuclear matter density and $Q^2$ increase.
This behavior is followed by the $\rho^+$ meson electromagnetic
form factors, $F_1^{* \rho}(Q^2)$,
$F_2^{* \rho} (Q^2)$, and $F_3^{* \rho} (Q^2)$,
giving a decrease of these $\rho$ meson form factors as the
nuclear matter density and $Q^2$ increase,
as illustrated in Fig.~\ref{fig4}(a)-(c).

In addition, we have predicted the in-medium $\rho^+$ meson charge radius,
that increases as the nuclear matter density increases.
Our prediction for the in-medium charge radius is
consistent with that obtained in Ref.~\cite{deMelo:2018hfw}, although different in magnitude, as
indicated in Fig.~\ref{fig5}(d).

Finally, we have predicted the $\rho^+$ meson magnetic and quadrupole moments in symmetric nuclear matter. We have found that the $\mu_\rho^*$ decreases
as the nuclear matter density increases.
The trends of our results are consistent with those obtained in Ref.~\cite{deMelo:2018hfw},
but different in magnitude. For the quadrupole moment, we have found that the
$\mathcal{Q}_\rho^*$ increases as the nuclear matter density increases.
Our predictions are consistent with those obtained in
Ref.~\cite{deMelo:2018hfw}, however, again, different in magnitude, in particular at higher
nuclear matter densities.

Overall, the predictions presented in this work are potentially testable by the
future experiments, such as JPARC E16~\cite{Aoki:2015qla}, and CLAS at
JLAB~\cite{CLAS:2008jqp,CLAS:2007dll}, as well as lattice
QCD~\cite{Hedditch:2007ex,Owen:2015gva,Wang:2025hew}.
For future work, we plan to apply the present approach to investigate the electromagnetic
form factors of other spin-1 mesons in a nuclear medium,
as well as pseudoscalar $D$ and $B$ mesons with heavy $c$ and $b$ quark contents,
where such studies may require
significant challenges, since heavy quark symmetry differs
fundamentally from the light quark symmetry. Furthermore, we hope that the present study
can provide useful guidance for lattice QCD in computing the electromagnetic form factors of the
$\rho$ meson in free space as well as in a nuclear medium.

\section*{Acknowledgments}
 This work was supported 
 by the PUTI Q1 Research Grant from the University of Indonesia (UI) under contract No. NKB
442/UN2.RST/HKP.05.00/2024. K.T.~was supported by Conselho Nacional de
Desenvolvimento Cient\'{i}fico e Tecnol\'ogico (CNPq, Brazil), Processes No. 304199/2022-2, and
FAPESP Process No.~2023/07313-6, and his work was also part of the projects, Instituto Nacional de
Ci\^{e}ncia e Tecnologia--Nuclear Physics and Applications (INCT-FNA), Brazil, Process
No.~464898/2014-5.

\section*{Data availability}
The data are not publicly available. The data are available from the authors upon reasonable request.
 
\bibliographystyle{elsarticle-num}
\bibliography{main}

\end{document}